\documentclass[10pt,journal,compsoc]{IEEEtran}
\usepackage{graphicx}
\usepackage{subfigure}
\usepackage{amsmath}
\usepackage{multirow}

\usepackage{amsfonts}
\usepackage{algpseudocode}
\usepackage{booktabs}
\usepackage{algorithm}
\usepackage{array}
\usepackage{color}
\usepackage{caption}
\usepackage{bbm}
\usepackage{ulem}

\newtheorem{definition}{Definition}
\newtheorem{definition1}{Problem Definition}

\floatname{algorithm}{Algorithm}

\ifCLASSOPTIONcompsoc
  \usepackage[nocompress]{cite}
\else
  \usepackage{cite}
\fi

\ifCLASSINFOpdf

\else

\fi

\begin{document} 
\title{Bayes-enhanced Multi-view Attention Networks for Robust POI Recommendation}

\author{\IEEEauthorblockN{Jiangnan Xia\IEEEauthorrefmark{1}, Yu Yang\IEEEauthorrefmark{1}, Senzhang Wang\IEEEauthorrefmark{2}, Hongzhi Yin, Jiannong Cao,~\IEEEmembership{Fellow,~IEEE}, \\
Philip S. Yu,~\IEEEmembership{Fellow,~IEEE}}
\IEEEcompsocitemizethanks{

\IEEEcompsocthanksitem Jiangnan Xia is with the School of Computer Science and Engineering, Central South University, Changsha, China, and also the Research Institute for Artificial Intelligence of Things, The Hong Kong Polytechnic University, Hong Kong, China, E-mail: 214712241@csu.edu.cn  

\IEEEcompsocthanksitem Senzhang Wang is with the School of Computer Science and Engineering, Central South University, Changsha, China. E-mail: szwang@csu.edu.cn

\IEEEcompsocthanksitem Yu Yang and Jiannong Cao are with the Research Institute for Artificial Intelligence of Things and the Department of Computing, The Hong Kong Polytechnic University, Hong Kong, China. E-mail: \{cs-yu.yang, jiannong.cao\}@polyu.edu.hk

\IEEEcompsocthanksitem Hongzhi Yin is with the School of Electrical Engineering and Computer Science, The University of Queensland, Australia. E-mail: h.yin1@uq.edu.au

\IEEEcompsocthanksitem P.S. Yu is with the Department of Computer Science, University of Illinois at Chicago, Chicago, USA. E-mail: psyu@uic.edu
}

% <-this % stops an unwanted space
\thanks{* Both authors contributed equally to this research. \dag~Corresponding author.}

}
%\thanks{Senzhang Wang is the corresponding author.}}

%{Shell \MakeLowercase{\textit{et al.}}: Bare Demo of IEEEtran.cls for Computer Society Journals}

\IEEEtitleabstractindextext{%
\begin{abstract}
POI recommendation is practically important to facilitate various Location-Based Social Network (LBSN) services, and has attracted rising research attention recently. Existing works generally assume the available POI check-ins reported by users are the ground-truth depiction of user behaviors. However, in real application scenarios, the check-in data can be rather unreliable (e.g. sparse, incomplete and inaccurate) due to both subjective and objective causes including positioning error and user privacy concerns. The data uncertainty issue may lead to significant negative impacts on the performance of the POI recommendation, but is not fully explored by existing works. To this end, this paper investigates a novel problem of robust POI recommendation by considering the uncertainty factors of the user check-ins, and proposes a Bayes-enhanced Multi-view Attention Network (BayMAN for short) to effectively address it. Specifically, we construct three POI graphs to comprehensively model the dependencies among the POIs from different views, including the personal POI transition graph, the semantic-based POI graph and distance-based POI graph. As the personal POI transition graph is usually sparse and sensitive to noise, we design a Bayes-enhanced spatial dependency learning module for data augmentation from the local view. A Bayesian posterior guided graph augmentation approach is adopted to generate a new graph with collaborative signals to increase the data diversity. Then both the original and the augmented graphs are used for POI representation learning to counteract the data uncertainty issue. Next, the POI representations of the three view graphs are input into the proposed multi-view attention-based user preference learning module. By incorporating the semantic and distance correlations of POIs, the user preference can be effectively refined and finally robust recommendation results are achieved. We conduct extensive experiments over three real-world LSBN datasets. The results show that BayMAN significantly outperforms the state-of-the-art methods in POI recommendation when the available check-ins are incomplete and noisy.

\end{abstract}

}

\maketitle

\IEEEdisplaynontitleabstractindextext
\IEEEpeerreviewmaketitle

\IEEEraisesectionheading{\section{Introduction}\label{sec:introduction}}

\IEEEPARstart Location-based social networks (LBSNs) such as Foursquare and Gowalla have become increasingly popular nowadays, where users can share their experience and check-in locations with their friends in real time. Point-of-Interest (POI) recommendation aims to recommend the POIs that the users will most likely visit next based on their historical POI check-in sequences. The accurate POI recommendation is practically important to support many LBSN services. For instance, the ride-hailing service can better predict the customers' next pick-up or drop-off locations based on the recommended POIs \cite{yao2018deep, fan2019golang}; travel Apps can use the recommended POIs to better plan the travel routes for customers \cite{taylor2018travel}. 

\begin{figure}[!t] \centering
    \subfigure[Incomplete check-ins] { 
            \includegraphics[scale=0.25]{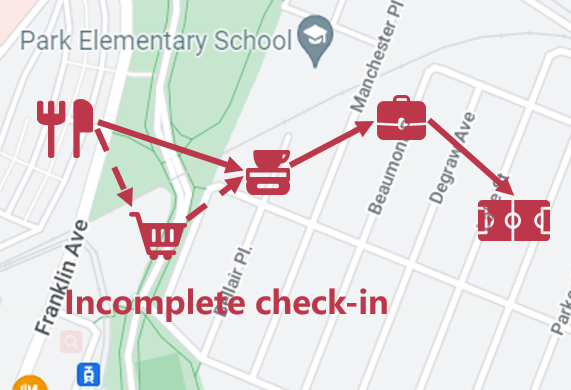}
            \label{introduction-into1_missed_check-ins} 
	}
	\subfigure[Incorrect check-ins] {
            \includegraphics[scale=0.25]{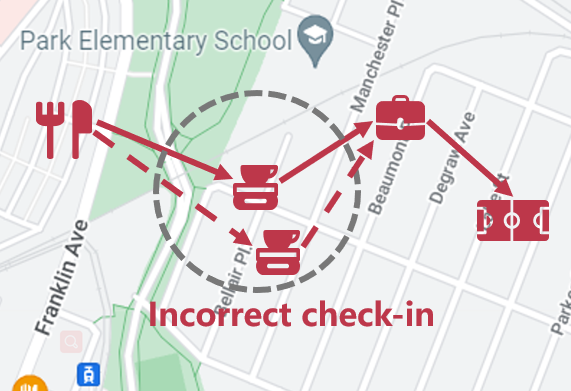}
            \label{introduction-into1_incorrect_check-ins} 
	}      
	\caption{Illustration of the unreliable POI check-ins.} %  (a) shows the incomplete check-in, and (b) shows the incorrect check-in in a check-in sequence.
	\label{introduction-into1} 
\end{figure}

Conventional POI recommendation methods can be generally categorized into collaborative filtering based methods such as matrix factorization~\cite{su2020fgcrec,rahmani2020joint,salakhutdinov2008bayesian} and sequence modeling methods such as Markov chain~\cite{xiong2020point, rendle2010factorizing}. One major issue of these approaches is that they heavily rely on feature engineering. Extracting informative features usually requires strong domain knowledge, which is inefficient and costly. 
Recently, deep learning models such as Recurrent Neural Networks (RNN) have been widely adopted in POI recommendation~\cite{yang2020location,chen2020next,wang2020next,zhao2020discovering} due to their strong capacity to learn the sequential dependencies from the check-in data \cite{liu2021attention, li2020group}. 
Some recent deep models can further capture both the spatial and temporal correlations among POI check-ins~\cite{liu2016predicting, zhao2020go, sun2020go, wang2020next, luo2021stan, yu2023self}. 
However, one limitation of existing methods is that they generally assume the available check-in data as a ground-truth description of the user behavior, which may not hold true in real application scenarios. In practice, check-in data may be unreliable due to both subjective and objective causes including positioning errors and user privacy concerns, and thus the data can be sparse, incomplete and full of noise~\cite{zhang2020modeling, sun2021point}.
Figure~\ref{introduction-into1} illustrates the incomplete and incorrect check-ins in users' check-in POI sequences. The dashed arrows indicate the ground truth POI transitions of a user, while the solid arrows represent the collected ones. The incorrect check-ins can be caused by positioning errors and introduce noise into the data. The incomplete check-ins can be due to users' privacy concerns, and thus make the data sparsity issue even worse. Both cases cause significant negative impacts on the POI recommendation and make this problem become challenging.

Recently, considerable research attention has been paid to addressing the check-in data sparsity issue in POI recommendations. 
Yin et al. \cite{yin2017spatial} proposed to address the sparsity issue by social regularization and spatial smoothing, which utilized both the collective preferences of the public in the target region and the personal preferences of users in the adjacent regions. However, it simply models the spatial correlations among the cell regions, but overlooks the global and personal semantic correlations among the POIs.
Zhao et al. \cite{zhao2020go} took the POI context information (e.g. geographical and user social influence) into consideration and jointly performed the POI context prediction and the next POI recommendation, which did not consider the global semantic correlations of the POIs. Chen et al. \cite{chen2022irlm} proposed to capture the global geographical influences and adopted an attention neural network for POI recommendation. 
\cite{wanggraph} proposed the GSTN model to learn the distance- and the global transition-based POI dependencies to improve the recommendation results. Both works ignore the semantic dependencies among the POIs of the personal check-ins. 

To sum up, although considerable efforts have been made on this problem, existing works still have the following shortcomings. First, current methods generally require external information such as context features and social relations of users to help address the data sparsity issue. They may not work well when such external information is unavailable. Thus a new method that can effectively address the data sparsity issue without relying much on external information is required. Second, although some attempts \cite{yin2017spatial,wanggraph} have been made to fuse the global POI semantics with the personal POI check-ins, there still lacks a unified framework to comprehensively capture and jointly fuse the semantic based POI correlations, the geographical distance based POI correlations and the personal POI preference. Third, as we discussed before, the observed POI check-in data can also be noisy and contain some incorrect check-ins. How to address the data uncertain issue, i.e., sparse and noisy, has not been fully studied by existing works and remains an open research issue.

To address the above issues, this paper proposes a Bayes-enhanced Multi-View Attention Networks (BayMAN) for robust POI recommendation, which consists of a Multi-view POI graph construction module, a Bayes-enhanced spatial dependency learning module and a multi-view attention based user preference learning module. Specifically, we first construct the following three POI graphs to comprehensively model the correlations among the POIs. The personal POI transition graph is constructed based on the historical check-in sequences of each user to reflect her/his POI preference. The semantic-based POI graph is constructed by aggregating the POI check-in sequences of all the users to model the semantic dependencies of the POIs. The distance-based POI graph is constructed to model the spatial closeness among POIs. As the personal POI transition graph is usually sparse and sensitive to noise, we design a Bayes-enhanced spatial dependency learning module for data augmentation on the personal POI transition graph. We propose a collaborative filtering-based graph sampling strategy to generate an augmented graph guided by the Bayesian posterior. Then both the original and the augmented graphs are used for POI representation learning to increase the data diversity and counteract the data uncertainty issue. To effectively learn the spatio-temporal POI preferences of users, we next propose a multi-view attention based user preference learning module. It integrates the POI representations learned from the views of semantic- and distance-based POI graphs into each user's personal POI preference representation, thus achieving a more robust result. Our contributions are summarized as follows.
\begin{itemize}
\item To the best of our knowledge, we are the first to study the novel problem of robust POI recommendation with the check-in data containing various uncertainty factors, and propose a Bayes-enhanced Multi-View Attention Network BayMAN to effectively address it.
\item From the personal view,  a Bayesian posterior guided graph augmentation approach is designed to generate a new personal POI transition graph with the collaborative signals from like-minded users. Then both the original and the augmented personal POI transition graphs are used for a more robust POI representation learning.
\item From the global view, a multi-view attention network is proposed to more comprehensively learn user preferences on POIs with the help of both semantic- and distance-based POI correlation graphs.
\item We conduct extensive experiments on three real-world LBSN datasets, i.e., Foursquare, NYC, and Gowalla. The results show that BayMAN significantly outperforms the state-of-the-art baseline methods over both the raw check-in data and the disturbed (e.g. incomplete and noisy) data, which verifies the effectiveness and robustness of BayMAN in POI recommendation.
\end{itemize}

% The remainder of this paper is organized as follows. Section \ref{section:2} provides a comprehensive literature review. We give a formal problem definition in Section \ref{section:3}, followed by presenting the details of the BayMAN in Section \ref{section:4}. Experimental results are presented in Section \ref{section:5} before we finally conclude the paper in Section \ref{section:6}.

\section{Related Work}
\label{section:2}
%The next POI recommendation is an emerging task, which refers to the process of suggesting the next possible POI for a user to visit based on his/her historical check-in sequence. The goal of this recommendation is to provide personalized and relevant suggestions to the user, enhancing their experience and helping them discover new places and activities that align with their interests. 
Existing studies on POI recommendation can be generally categorized into conventional POI recommendation methods and deep learning based methods. In the following sections, we will review related works from the two aspects.

\subsection{Conventional POI Recommendation Methods}
Conventional POI recommendation methods mainly contain collaborative filtering and sequence modeling methods, such as matrix factorization (MF) \cite{salakhutdinov2008bayesian} and Markov chain (MC) \cite{rendle2010factorizing}. MF based approaches factorize the user-POI matrix into a low-rank matrix representation that captures the underlying characteristics of users and POIs \cite{koren2009matrix}. Although MF can effectively identify the latent factors, it is not capable of capturing the temporal dependencies of the check-in sequences. In addition, MF is computationally expensive when decomposing large matrices. MC is also widely used in POI recommendations due to its exceptional ability to capture the continuous movement patterns of users \cite{rendle2010factorizing}. The idea of MC based models is to model the user's historical check-in sequence as a Markov chain, and leverage the estimated POI transition probabilities to recommend the next POI. However, these methods have limited modeling capabilities as they merely model the transitions between POIs, yet explore the complex spatial and temporal dependencies from users' check-in behaviors.

\subsection{Deep Learning-based POI Recommendation}
Recently, deep learning based methods have attracted considerable research attention and demonstrated remarkable performance \cite{islam2022survey}, due to their effectiveness in modeling nonlinear and complex user-item interactions. RNN based models \cite{yang2020location,chen2020next} were first adopted to POI recommendations because of their effectiveness in learning the sequential dependencies from users' check-in sequences.
Previous researches have demonstrated that explicit spatial and temporal information can greatly enhance the recommendation performance \cite{liu2016predicting}. An early work, STRNN \cite{liu2016predicting}, combines RNN units with temporal and spatial matrices, incorporating spatial and temporal distance between consecutive visits in linear interpolation and boosting the recommendation performance. Time-LSTM \cite{zhu2017next} proposes to add time gates to the LSTM to capture the time lags between consecutive items for sequential recommendation. Inspired by Time-LSTM, Zhao et al. \cite{zhao2020go} further propose STGN which integrates the dedicated time and distance gates into the LSTM, explicitly modeling the spatio-temporal intervals between consecutive check-ins 
for the next POI recommendation. 
However, the above methods merely model the spatial and temporal dependency between consecutive check-ins. 

To explore the dependencies between discontinuous check-ins, Sun et al. \cite{sun2020go} propose LSTPM that uses a geo-dilated RNN to fully exploit the geographical relations among non-consecutive POIs and a context-aware nonlocal network structure to explore the spatio-temporal correlations between historical and current trajectories. \cite{wang2021attentive} proposes ASGNN which constructs a user-POI interaction graph for each user to explore the semantic relationships between POIs from a personal view. Luo et al. \cite{luo2021stan} propose STAN that adopts an attention architecture to aggregate the spatio-temporal correlation of non-adjacent locations and non-consecutive visits within users’ trajectories. However, the above approaches fail to fully explore the relationships between POIs from a global view.
Although Wang et al. \cite{wanggraph} propose GSTN which leverages the global semantic correlations and distance dependencies between POIs to improve the recommendation performance, GSTN does not learn the semantic dependencies among POIs from a personal view.
% Recently, Wang et al. \cite{wanggraph} proposes GSTN which explored the semantic correlations and distance dependencies between POIs from a global view to improve the recommendation performance, but they neglect the learning of semantic dependencies among the POIs of the personal check-ins.

Different from existing works, we aim to simultaneously explore the semantic correlations among the POIs from both the global and personal views, as well as their geographical distance dependencies. Meanwhile, existing approaches generally regard the check-in data as a ground-truth description of user behavior, which may not hold true in real application scenarios. In this paper, we study the novel problem of robust POI recommendation when the given POI check-ins of users are unreliable. %In real application scenarios, the check-in data may be incomplete and inaccurate due to various reasons, posting great challenges for existing methods. This paper aims to propose a more robust recommendation algorithm that can alleviate both data sparsity and inaccuracy issues.

\section{Problem Statement}
\label{section:3}
We will first define some terminologies to help state the studied problem, followed by a formal problem definition.
\begin{definition}
\textbf{Check-in sequence.} We denote the user set by $\mathcal{U}=\{u_{1},u_{2},...,u_{M}\}$ and the POIs set by $\mathcal{V}=\{v_{1},v_{2},...,v_{N}\}$, where $M$ is the number of users and $N$ is the number of POIs. Each POI is associated with its longitude and latitude. Each user's POI check-in sequence 
% $s_u=\{v_i^{t_i} \mid i=1,2,...,n\}$ 
$s_u=\{v_{t_i} \mid i=1,2,...,n\}$ is the POI transition histories of the user $u\in \mathcal{U}$, where $n$ is the number of check-ins of the user and $v_{t_i}$ represents the visited POI at time $t_i$.
\end{definition}
Note that user's check-in sequence $s_u$ may be incomplete and noisy due to some missing and incorrect check-ins.

\begin{definition}
\label{def:2}
 \textbf{Personal POI transition graph.} The personal POI transition graph of user $u\in \mathcal{U}$ is defined as a directed graph $G_u = (V_u,E_u,A_u)$, where $V_u$ is a set of POIs that $u$ visited. The edge $e_{u}^{i,j}\in E_u$ indicates user $u$ visited POI $v_j$ after $v_i$. $a_{u}^{i,j}\in A_u$ is the edge weight of $e_{u}^{i,j}$ measured by the probability of visiting $v_j$ after $v_i$ within $u$'s POI transitions $s_u$.
 \end{definition}
 \begin{definition}
% % Preference-based POI graph
\label{def:3}
 \textbf{Semantic-based POI graph.} The semantic-based POI graph is defined as a directed graph $G_S = (\mathcal{V},E_S,A_S)$, where the nodes $\mathcal{V}$ are POIs and the edge $e_{S}^{i,j}\in E_S$ represents any user visited POI $v_j$ after $v_i$. $a_{S}^{i,j}\in A_S$ is the edge weight of $e_{S}^{i,j}$ measured by the overall probability of visiting $v_j$ after $v_i$ across all users.
 \end{definition}
 \begin{definition}
 \label{def:4}
 \textbf{Distance-based POI graph.} The distance-based POI graph $G_D=(\mathcal{V}, E_D, A_D)$ is defined as an undirected graph, where the nodes $\mathcal{V}$ are POIs. An edge $e_{D}^{i,j}\in E_D$ reflects the distance-based relations
 % exists between $v_j$ and $v_i$ if any user has consecutive check-in 
 between $v_j$ and $v_i$. We employ the Euclidean distance between $v_i$ and $v_j$ as the edge weight $a_{D}^{i,j}\in A_D$ associated with the edge $e_{D}^{i,j}$ to measure their distance-based closeness.
 \end{definition}
Based on the above defined terminologies, we present a formal problem definition as follows. 
\begin{definition1}
Given a set of users $\mathcal{U}=\{u_1,u_2,...,u_M\}$, a set of POIs $\mathcal{V}=\{v_1,v_2,...,v_N\}$, and each user's unreliable historical check-in sequences $S=\{s_u \mid u\in\mathcal{U}\}$ where $s_u=\{v_{t_i} \mid i=1,2,...,n\}$, our objective is to estimate each user’s preference based on the historical check-in sequence and recommend top-$k$ POIs to each user in descending order of his/her next check-in probability.
%in descending order of their access probability.
% we aim to consider the historical POI visits $\{l_{t_1}, l_{t_2},...,l_{t_i}\}$ and user $u$ to recommend the most interesting POI $l_{t_{i+1}}$ from $\mathcal{L}$ for user $u$ at the next timestamp.
\end{definition1}

\begin{figure*}[!t]
    \centering
    \includegraphics[scale=0.38 ]{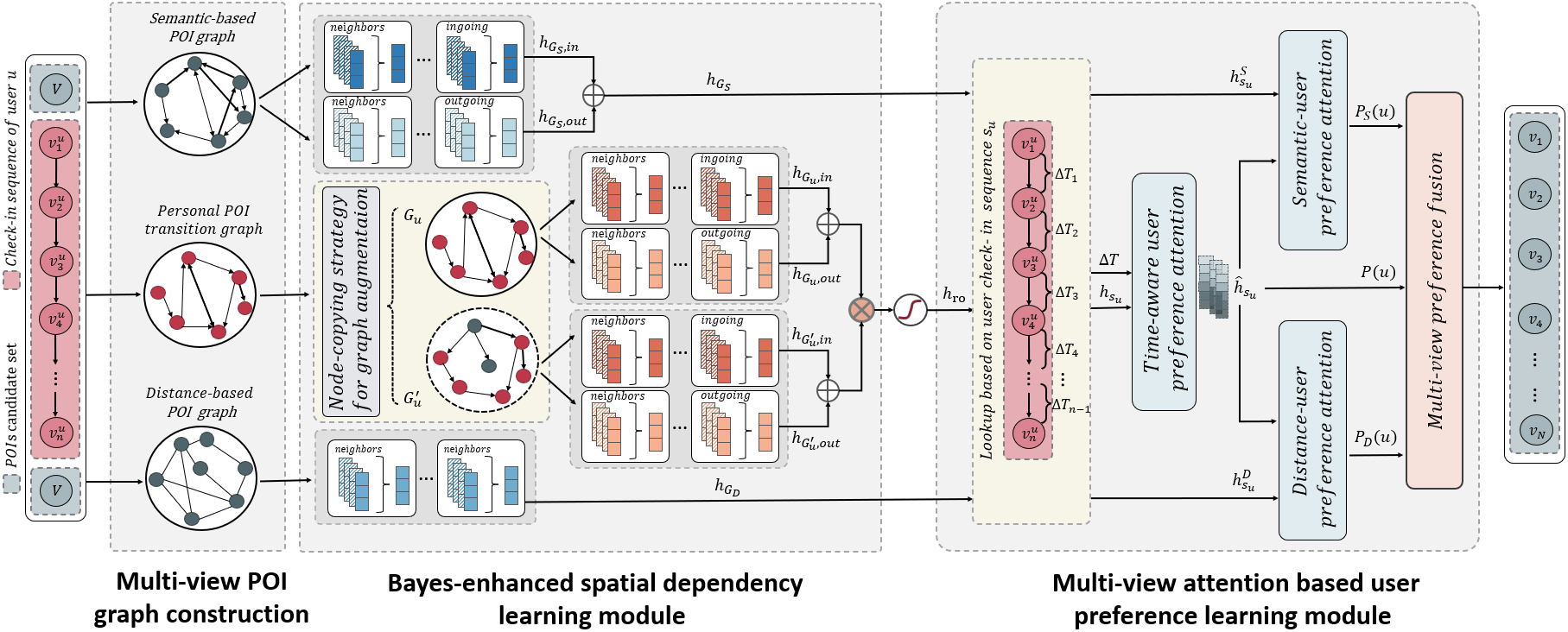}  %frame1.png}
    \caption{The overall framework of the proposed BayMAN model.}
    % \caption{The overall framework of the proposed BayMAN model. BayMAN consists of a Bayes-enhanced spatial dependency learning module and a multi-view preference attention module.}
    \label{framework}
\end{figure*}

\section{Methodology}
\label{section:4}
%In this section, we will first introduce the overall framework of BayMAN, and then present the technical details.

\subsection{Model Framework}
Figure~\ref{framework} illustrates the overall framework of BayMAN, which consists of multi-view POI graph construction, Bayes-enhanced spatial dependency learning module and multi-view attention based user preference learning module. We construct three views of POI graphs as defined in Section \ref{section:3} to model the personal POI transition relations, the semantic dependencies as well as the geographical closeness of POIs, respectively. The details will be introduced in Section~\ref{multi-view graph construction}. To alleviate the uncertainty of the personal POI check-in sequences, a novel Bayes-enhanced data augmentation approach is then designed to generate an augmented POI transition graph based on Bayesian posterior as shown in the central part of Fig. \ref{framework}. We learn the spatial dependencies among POIs from both the original and the augmented personal POI transition graphs and fuse them together to obtain robust POI representations from the personal view.
Meanwhile, we embed the semantic- and distance-based POI graphs to obtain the POI representations from global views. The details will be presented in Section~\ref{bayes-enhance SD Learning}. To effectively learn the spatio-temporal POI preferences of each user, we propose a multi-view attention based user preference attention module which is shown in the right part of Figure~\ref{framework}. 
We first design a time-aware attention layer to embed the time lags between consecutive POI check-ins into the personal POI representations, successfully learning each user's personal POI preference. Then, we employ two attention layers to refine the user preference by incorporating the semantic- and distance-based global POI correlations, respectively. Next, we fuse the personal POI preference, the semantic-aware global POI preference and the distance-aware global POI preference. We will elaborate on this step in Section~\ref{multi-view preference attention}. Finally, we employ a softmax layer to predict the next POI for recommendations. The overall objective function will be described in Section~\ref{objective function}.
% \begin{figure*}[!t]
% \centering
%     \subfigure[The personal POI transition graph of $u$]{
%         \label{fig:PTG}
%         \includegraphics[width=0.29\textwidth]{image/personal-graph.png}}
%     \hspace{0.15in}
%     \subfigure[The semantic-based POI graph]{
%         \label{fig:GTG}
%         \includegraphics[width=0.29\textwidth]{image/sem-graph.png}}
%     \hspace{0.15in}
%     \subfigure[The distance-based POI Graph]{
%         \label{fig:DPG}
%         \includegraphics[width=0.29\textwidth]{image/dis-graph.png}}
%     % \hspace{0.1in}
%     \caption{Illustration of the multi-view POI graph construction.}
%     \label{fig:multi-view POI graph construction}
% \end{figure*}

\subsection{Multi-view POI Graph Construction}
\label{multi-view graph construction}
We construct three views of POI graphs to model the personal POI transition relations, the semantic dependencies and the geographical closeness among POIs, respectively.

\textbf{Personal POI transition graph construction.} The next likely visiting POI of a user is highly correlated with his/her previous visited POIs. To model the personal POI transition relations, we construct a personal POI transition graph for each $u\in\mathcal{U}$ as in Definition \ref{def:2}. Each edge $e_u^{i,j}\in E_u$ is associated with a edge weight $a_u^{i,j}\in A_u$ measuring the probability of visiting $v_j$ after $v_i$ within the POI check-in sequence $s_u$ as shown in Eq. (\ref{eq:weight_PTG})
\begin{equation}
a_u^{i,j} = \dfrac{freq_u(v_i,v_j)}{freq_u(v_i)},
\label{eq:weight_PTG}
\end{equation}
where $freq_u(v_i,v_j)$ is the number of the POI transition from $v_i$ to $v_j$ of $u$, and $freq_u(v_i)$ is the times that $u$ checks in at $v_i$. 
%Fig.~\ref{fig:PTG} shows an example of a personal POI transition graph.
Since the POI check-ins $s_u$ is unreliable, the personal POI transition graph $G_u$ is usually incomplete and inaccurate with noise.

\textbf{Semantic-based POI graph construction.} The collective POI visiting sequences of all the users can largely reflect the semantic correlations among POIs. For example, the check-in sequence \textit{meal-market} appears more frequently than the sequence \textit{meal-sports} in the check-in dataset, which means users prefer to go shopping after meal than go sports. To capture such common check-in sequence patterns, we construct a semantic-based POI graph $G_S = (\mathcal{V}, E_S, A_S)$ based on the check-in sequences $S=\{s_u \mid u\in\mathcal{U}\}$ of all the users as in Definition \ref{def:3}.
% users' common POI preference. 
%Fig.~\ref{fig:GTG} shows an example of the semantic based POI graph.
%The global POI transition graph $G_T = (\mathcal{V}, E_T, A_T)$ is directed, where the nodes are POIs in $\mathcal{V}$ and the edge $e_T^{i,j}\in E_T$ connecting source POI $v_i$ and target one $v_j$ indicates there exists at least a visit of POI $v_j$ after $v_i$.
Each edge $e_S^{i,j}$ is associated with a edge weight $a_S^{i,j}\in A_S$ measuring the probability of visiting $v_j$ after $v_i$ in the check-in sequences $S$ as shown in Eq. (\ref{eq:weight_GTG})
\begin{equation}
a_S^{i,j} = \dfrac{freq(v_i,v_j)}{freq(v_i)},
\label{eq:weight_GTG}
\end{equation}
where $freq(v_i,v_j)$ is the number of the POI transition from $v_i$ to $v_j$ in $S$, and $freq(v_i)$ is the times of visiting $v_i$. 

\textbf{Distance-based POI Graph Construction.} The first law of geography says that ``\textit{near things are more related than distant thing}'' \cite{tobler1970computer}. Existing works have shown that, in location-based social networks, a user's next check-in location is usually near to the current check-in location \cite{zhang2021interactive, zhao2020discovering}. To capture the geographical closeness among POIs, we further construct a distance-based POI graph $G_D=(\mathcal{V}, E_D, A_D)$ over the POIs set $\mathcal{V}=\{v_1,v_2,...,v_N\}$ based on Definition \ref{def:4}. 
%Fig~\ref{fig:DPG} shows an illustration of the distance-based POI graph. 
The edge weight $a_D^{i,j}\in A_D$ is calculated by Eq. (\ref{Deltad})
\begin{equation}
a_D^{i,j} = \left \{ \begin{array}{ccc}
\dfrac{1}{dis(v_i,v_j)} &, &0< dis(v_i,v_j)\le \Delta d\\
0 &,   &otherwise
\end{array}\right.
\label{Deltad}
\end{equation}
where $dis(v_i,v_j)$ represents the Haversine distance between $v_i$ and $v_j$ given their longitudes and latitudes. We employ the reciprocal of $dis(v_i,v_j)$ as the edge weight $a_D^{i,j}\in A_D$. The distance threshold $\Delta d$ is a hyper-parameter.

\subsection{Bayes-enhanced Spatial Dependency Learning}
\label{bayes-enhance SD Learning}
Since the personal POI transition graph is sparse and sensitive to noise, directly applying Graph Convolutional Network (GCN) cannot learn desirable POI representations.
We propose a novel Bayes-enhanced data augmentation approach to generate an augmented personal graph, and then employ GNN to learn robust POI representations from the original and augmented personal POI transition graphs, the semantic- and the distance-based POI graphs, respectively.
% In the Bayes-enhanced spatial dependency learning module, we perform representation learning on the constructed three graphs. Specifically, we adopt Graph Convolutional Network (GCN) on the semantic- and the distance-based POI graphs to learn the POI representations of the two views, respectively. For the personal POI transition graph, due to its sparsity and sensitivity to noise, directly applying GCN may not obtain desirable POI representations. To this end, we propose a novel Bayes-enhances data augmentation approach to generate an augmented personal graph, and perform representation learning on both the original and the augmented graphs to obtain robust POI representations.

\subsubsection{Personal POI Transition Graph Augmentation with Collaborative Filtering Strategies}

As the check-ins of a user are usually sparse, the personal POI transition graph is more sensitive to the issues of data uncertainty, i.e., incomplete and incorrect check-ins. Thus, relying on the original personal POI transition graph is not enough to learn a robust POI representation. To address this issue, we want to create an augmented graph that can pick up collaborative signals from like-minded users. To select beneficial collaborative signals, we calculate the Bayesian posterior based on the original and the like-minded user to guide the augmentation of the new graph. With the augmented personal graph as complementary information, the data uncertainty issue can be effectively alleviated \cite{sun2020framework}.

Collaborative filtering is a common technique in recommendations. It recommends similar items to users who have similar preferences \cite{he2017neural}. Some POIs that are missing or incorrect in the original user sequence may appear in the like-minded user’s sequence. Therefore, we want to find a like-minded user for the original user and capture the collaborative signals to support the augmented graph generation. We measure the similarity between two users $u_p$ and $u_q$ by analyzing the similarity between their historical check-in sequences based on the following formula,
\begin{equation}
sim(u_p, u_q) =   \frac{S(u_p) \cap S(u_q)}{S(u_p) \cup S(u_q)}
\label{user_sim}
\end{equation}
where $u_p$ is the original user. $S(\cdot)$ represents the set of the POIs in the user's historical check-in sequences. According to this formula, if users visit more same POIs, their POI preferences are more similar.

The most like-minded user $u_q \in \mathcal{U}$ letting $sim(u_p, u_q)$ reaches the maximum value. Although determining the $u_q$, we cannot directly utilize all of his/her information due to its large size and noise. Therefore, it is necessary but hard to select the collaborative signals that are beneficial to $u_p$. In fact, each user's different preferences cause the corresponding personal graph to follow a certain distribution, and the augmented graph containing collaborative signals should also satisfy this distribution. To this end, we extend the Bayesian paradigm to the task of the next POI recommendation, treating the original personal POI transition graph as a realization from a parametric family of random graphs. Following the Bayesian idea\cite{zhang2019bayesian}, we then make posterior inference $p(G_{u_p}^{\prime} \mid G_{u_p}, G_{u_q})$, where $G_{u_p}^{\prime}$, $G_{u_p}$ and $G_{u_q}$ are the augmented, the original and the like-minded personal graph.
Inspired by \cite{pal2019bayesian}, we propose a graph generation model based on node copying to generate an augmented graph. The resulting $G_{u_p}^{\prime}$ follows both the distribution of the original graph and contains the collaborative signals from the most like-minded user. Unlike \cite{pal2019bayesian}, which focuses on node classification tasks based on a single graph, our approach is designed for two personal graphs and further considers the temporal information in the sequence during the generation process.

\begin{figure}[t] \centering
	\includegraphics[scale=0.26]{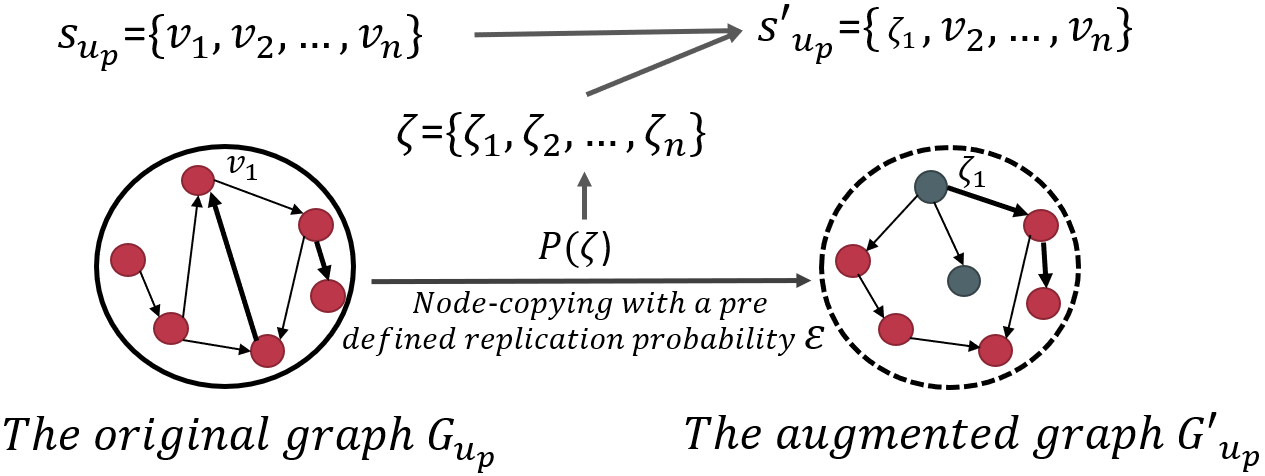}     
	\caption{Illustration of Bayes-enhanced data augmentation approach with node-copying strategy}
	\label{augmented} 
\end{figure}

The augmented graph is generated as shown in Fig. \ref{augmented}. We introduce an auxiliary random vector $\zeta=[\zeta_1,\zeta_2,...,\zeta_n]^T$ to facilitate the node copying operation for augmenting the graph, where $n$ is the sequence length of user $u_p$. We denote the adjacency matrixs of $p(G_{u_p}^{\prime}$, $G_{u_p}$ and $G_{u_q})$ as $A_{u_p}^{\prime}$, $A_{u_p}$ and $A_{u_q}$. Each entry $\zeta_i$ represents a POI node which means the row $i$ of $A_{u_p}$ will be replaced by the row $\zeta_i$ of $A_{u_q}$, resulting in a new graph $G_{u_p}^{\prime}$. In this way, the collaborative signals based on $u_q$ will be integrated into $G_{u_p}^{\prime}$. The entries in $\zeta$ are assumed to be independent of each other, and the distribution $p(\zeta)$ of $\zeta$ is proportional to a predefined similarity measurement, such as user similarity and POI similarity learned from the original personal POI transition graph $G_{u_p}$ and user check-in data $C$. In the POI recommendation, we need to learn a suitable replication distribution $p(\zeta)$ for each user to facilitate the augmentation of the personal POI transition graph.

The entries in $\zeta$ are assumed to be independent, and the distribution $p(\zeta)$ is proportional to a predefined similarity measurement, such as POI similarity learned from $G_{u_p}$ and $G_{u_q}$. In POI recommendation, we need to learn a suitable distribution $p(\zeta)$ for each user to guide the augmentation of $G_{u_p}$.

To measure the similarity between POIs %that belong to user $u_p$ and $u_q$
, we consider both the impact of the check-in time and the geographic distance. Thus the similarity between two POIs $v_i \in s_{u_p}$ and $v_j \in s_{u_q}$ can be calculated by the following formula,
% \begin{equation}
% sim(v_i, v_j)_v =  \frac{1}{dis(v_i,v_j) + \left| T_{v_i} - T_{v_j}\right| }  
% % dis(v_i,v_j) + \left| T_{v_i} - T_{v_j} \right|
% \end{equation}
\begin{equation}
\begin{split}
& sim(v_i, v_j)_v = \left\{
\begin{aligned}
 & 1  ,\qquad  if \quad v_i = v_j \quad and  \quad t({v_i}) = t({v_j}) \\
 & \sigma(\frac{1}{dis(v_i,v_j) + \left| t({v_i}) - t({v_j})\right| })  ,\quad else \\
\end{aligned}
\right.
\end{split}
\end{equation}
where $dis(\cdot, \cdot )$ represents the Haversine distance between the two POIs. $ t({v_i})$ and $t({v_j})$ indicate the visiting time of POIs $ v_i$ and $v_j$, respectively. $\sigma(\cdot)$ is a \textit{tanh} activation function. Two POIs are more similar if they have a closer geographic distance and are visited at similar times.

We then define the posterior distribution of $\zeta$ as follows
\begin{equation}
\begin{split}
%& u_q : max \{ sim(u_p, u_q)_u \}, \\
& p(\zeta \mid G_{u_p},G_{u_q} ) = \prod_{i=1}^{n} p(\zeta_i \mid G_{u_p}, G_{u_q}), \\
& p(\zeta_i= v_l \mid G_{u_p}, C) =  \frac{sim(v_i, v_l)_v}{\sum_{j=1}^{|s_{u_q}|} sim(v_i,v_j)_v},
\end{split}
\end{equation}
% 得到分布后，采样！！
where $G_{u_p}$ and $G_{u_q}$ represent the original and the like-minded personal graph. Note that the two graphs do not share the same weight matrix. And $v_i \in s_{u_p}$ and $v_l \in s_{u_q}$. $s_{u_p}$ and $s_{u_q}$ are the check-in sequences of users $u_p$ and $u_q$, respectively. 

After obtaining the distribution $p(\zeta)$, we replicate the $\zeta_i$ node of $G_{u_q}$ at the position of the $i$ node of $G_{u_p}^{\prime}$, and the corresponding process can be represented as $\mathbbm{1}\{G_{u_p, i}^{\prime} =G_{u_q, \zeta_i} \}$. Eventually, the generation model can be defined as
\begin{equation}
P(G_{u_p}^{\prime} \mid G_{u_p}, \zeta) = \prod_{i=1}^{n}  \varepsilon^{\mathbbm{1}\{G_{u_p, i}^{\prime} =G_{u_q, \zeta_i} \}} (1-\varepsilon)^{\mathbbm{1}\{G_{u_p, i}^{\prime} =G_{u_p, i} \}},
\label{retain}
\end{equation}
where $0 < \varepsilon \leq 1$ represents the probability of replication for each POI node. $1-\varepsilon$ represents the probability of retaining the original structure and the features. The augmented graph $G_{u_p}^{\prime}$ arguments the original personal graph $G_{u_p}$ with collaborative signals shown as new edges and nodes, which increase diverse connectivity among POIs and able to mitigate the unreliability issue in users' check-in sequences.

\subsubsection{Personal POI Transition Graph Representation Learning}
With the original and the augmented personal POI transition graphs, we next conduct graph representation learning on them separately and then fuse their embedding together to obtain robust POI representations.

We employ Graph Convolutional Network (GCN) to learn the POI representation on both graphs. Specifically, we conduct spectral graph convolutions \cite{kipf2017semi} as follows
\begin{equation}
    h_G = f(H, A) = ReLU(\bar{D}^{-\frac{1}{2}}\widetilde{A}\bar{D}^{-\frac{1}{2}}HW),
    \label{GCN}
\end{equation}
where $f(\cdot,\cdot)$ represents the GCN operation, $h_G$ means the feature vector representation learned by the GCN layer, and $H$ and $A$ are the node embedding and the adjacency matrix of the graph $G$, respectively. $\widetilde{A}$ is $A$ with added self-connections. $D$ is the degree matrix and $W$ represents the learnable weight matrix. 
$ReLU(\cdot)$ is a rectified linear unit which is a nonlinear activation function.
The above equation can be interpreted as the first-order approximation of the local spectral filtering network, which itself is the local approximation of the spectral network convolved in the frequency domain using the graph Fourier transform according to the convolution theorem \cite{defferrard2016convolutional}.

In general, graph convolution operation reflects the correlations between POIs in the graph. In practice, two POIs show a strong correlation with each other if they often co-occur in the same user's check-in sequence. For example, if the user is used to go shopping after meals, he/she is more likely to check in at a nearby supermarket after meals. Such complex and highly non-linear transition dependencies can be captured by the adjacency matrix in Eq. (\ref{GCN}).
% , so that the hidden features of one node (POI) can be propagated to its neighbor nodes. 

The representations of the original and the augmented personal graphs can be learned as follows,
% \mathcal
\begin{equation}
    \begin{split}
        \begin{aligned}
        & h_{G_{u,in}}  = GCN({G}_{u}, A_{u,in}), \\
        & h_{G_{u,out}} = GCN({G}_{u}, A_{u,out}),\\
        & h_{G_{u,in}^{\prime}} = GCN({G}_{u}^{\prime}, A_{u,in}^{\prime}), \\
        & h_{G_{u,out}^{\prime}} = GCN({G}_{u}^{\prime}, A_{u,out}^{\prime}),
        \end{aligned}
    \end{split}
    \label{gcn_op}
\end{equation} 
where ${G}_{u}$ is the personal POI transition graph and $G_{u}^{\prime}$ is the augmented graph. $A_{u,in}, A_{u,in}^{\prime}, A_{u,out}$ and $A_{u,out}^{\prime}$ are in- and out-going adjacency matrices of the original and augmented graphs, respectively. $h_{G_{u,in}}, h_{G_{u,in}^{\prime}}, h_{G_{u,out}}$ and $h_{G_{u,out}^{\prime}}$ are the corresponding POI representations.

Then, we fuse the learned representations of the original and the augmented graphs. We employ a multi-channel fusion layer described in Eq. (\ref{f-fuse-global}) to integrate the ingoing and outgoing POI representations,
\begin{equation}
    \begin{split}
        \begin{aligned}
        & h_{G_u} = W_{u_1} \odot h_{G_{u,in}} + W_{u_2} \odot h_{G_{u,out}} \\
        & h_{G_{u}^{\prime}} = W_{u_1}^{\prime} \odot h_{G_{u,in}^{\prime}} + W_{u_2}^{\prime} \odot h_{G_{u,out}^{\prime}} 
        \end{aligned}
    \end{split}
    \label{f-fuse-global}
\end{equation}
where $\odot$ represents the Hadamard product. $W_{u_1}$, $W_{u_2}$, $W_{u_1}^{\prime}$ and $W_{u_2}^{\prime}$ are the corresponding trainable parameters, which reflect the importance of ingoing and outgoing representations. Then, we combine $h_{G_u}$ and $h_{G_{u}^{\prime}}$ as follows.
\begin{equation}
    h_{ro} = \sigma(h_{G_u} || h_{G_{u}^{\prime}})
    \label{f-fuse}
\end{equation}
where $\sigma(\cdot)$ is a \textit{tanh} activation function and $||$ represents the concatenation operation.

\subsubsection{Semantic- and Distance-based POI Graphs Representation Learning}
%Meanwhile, we need to consider complex semantic relations and geographic closeness among POIs. In practice, two POIs show a strong correlation with each other if they often co-occur, or has close proximity. For instance, if most users tend to shop after meals, a user is likely to check-in a nearby supermarket after she/he has dined. 
For the semantic- and distance-based POI graphs, we again employ GCN to embed the semantic correlations and the geographic closeness among the POIs as their representations. Similar to Eq.(\ref{gcn_op}) and Eq.(\ref{f-fuse-global}), the POI representation learning over the semantic- and distance-based POI graphs is conducted through the following formulas,
\begin{equation}
    \begin{split}
        \begin{aligned}
        & h_{G_D} = GCN({G}_{D}, A_{D}),\\
        & h_{G_{S,in}}  = GCN({G}_{S}, A_{S,in}), \\
        & h_{G_{S,out}} = GCN({G}_{S}, A_{S,out})\\
        & h_{G_S} = W_{S_1} \odot h_{G_{S,in}} + W_{S_2} \odot h_{G_{S,out}}
        \end{aligned}
    \end{split}
    \label{SD-gcn-op}
\end{equation}
where $h_{G_D}$ and $h_{G_S}$ are the POI representations learned from the semantic- and the distance-based POI graphs, respectively.

% 用户偏好模块
\subsection{Multi-view Attentions for User Preference Learning}
\label{multi-view preference attention}
To effectively learn the spatio-temporal POI preference of each user, we propose a multi-view attention based user preference learning module to comprehensively fuse three types of POI preferences learned from the personal POI transition graph, the semantic- and distance-based POI graphs. Specifically, a time-aware user preference attention layer is designed to learn the sequential and temporal POI preference from the representation of the personal POI check-in sequence.
Then, we design two attention layers to refine the user preference by learning semantic- and distance-based POIs correlations from the semantic- and distance-based POI graphs, respectively. Lastly, we fuse the preferences learned from the above three attention layers to fully explore users' POI preferences from different views. %We detail the corresponding process in this section.

\subsubsection{Time-aware Attention for Learning Personal User Preference} 
% time interval picture
% \begin{figure}[!t] \centering
%     \subfigure[Check-ins in one hour] { 
% 		\includegraphics[scale=0.21]{image/1hour.png}     
% 	}
% 	% \hspace{1.5cm}
% 	\subfigure[Check-ins in ten hours] {
% 		\includegraphics[scale=0.21]{image/10hour.png} 
% 	}     
% 	\caption{Illustration of two check-in sequences with different time lags}
% 	\label{T-atten} 
% \end{figure}
We first learn the temporal dependencies from the user check-in sequence to capture the user preference. A novel attention layer is proposed to introduce time lag into the attention mechanism, which can learn different weights for each POI in the trajectory sequence and consider the influence of check-in time lags on user preference.

As attention mechanisms can effectively capture the internal correlations of the sequence data, it has been widely used in various sequence modeling problems \cite{vaswani2017attention}. However, the conventional attention layer does not consider the time lag in the input sequence, which is critical for modeling the user's temporal behavior \cite{li2020time}. For example, two users have the same historical check-in sequence. One completes these check-ins within ten hours, while the other completes them in only one hour. In this case, their preferences may be significantly different, because the correlation between two check-ins is higher if they happen in a short time lag than that in a long time lag \cite{zhu2017next, wanggraph}. Therefore, the time lags between check-ins should not be ignored when learning the sequential dependency of the check-in sequence.

In this light, we integrate explicit time lags into the interaction. Given the user check-in sequence $s_{u} = \{v_{1}, v_{2},...,v_{n}\}$, the trajectory representation $h_{s_{u}}$ is selected from $h_{ro}$ based on $s_{u}$. 
$\Delta T $ is a vector that contains time lags that are calculated based on $s_{u}$, in which $\Delta T_{i} = t(v_{i+1}) - t(v_{i})$. We then design a time-aware attention layer to encode the representation of the user's POI check-in sequence by Eq. (\ref{time attention}),
\begin{equation}
    \begin{split}
        \begin{aligned}
        & Q, K, V = W_Q h_{s_{u}}, W_K h_{s_{u}}, W_V h_{s_{u}} \\
        % & \hat{h}_{s_{u}} = Time Attention(Q,K,V, \Delta T)\\
        & \hat{h}_{s_{u}} = softmax(\frac{Q K^T + \Delta T}{\sqrt{d}} )V \\  %+ \varepsilon
        & P(u) = \sum_{i=1}^n \hat{h}_{s_{u},i}
        \end{aligned}
    \end{split}
    \label{time attention}
\end{equation}
where $h_{s_{u}} \in R^{n\times c}$ is the embedding of the user's POI check-in sequence. $W_i$ is the trainable weight matrix. $\hat{h}_{s_{u}} $ denotes the learned personal trajectory representation. $P(u)$ is the learned personal preference of users, where $\hat{h}_{s_{u},i}$ is the $i$-th row in $\hat{h}_{s_{u}}$ which is the representation of users' preference to the node $v_{i}$ in user's sequence $s_u$. %visited by the user in the corresponding trajectory.
Notably, the learning of $P(u)$ only leverages the representations learned from the original and augmented personal POI transition graphs, which does not consider the semantic dependencies and geographic closeness of the POIs.

\subsubsection{User Preference Refinement by Incorporating Multi-view POIs Correlations}
\label{P(u)}
% 为了更加全面的了解用户偏好, 从而进一步抑制个人签到序列中的的不确定性问题，我们分别利用从语义图和距离图学到的两类POI表示来补充个人偏好
To more comprehensively and robustly capture user preferences, we further refine the learned user preference with the semantic dependencies and geographic closeness of the POIs. To this aim, we design two preference refinement attention layers. They are a semantic-user preference attention to learn the semantic correlations of POIs from the semantic-based POI graph and a distance-user preference attention to embed the geographic closeness among POIs in the distance-based POI graph.

Specifically, for semantic-user preference attention, we first select the semantic trajectory representation $h_{s_{u}}^S$ from $h_{G_{S}}$ based on the user sequence $s_{u}$. Then, the attention mechanism is used to fuse the personal trajectory representation $\hat{h}_{s_{u}}$ with the semantic trajectory representation $h_{s_{u}}^S$. By incorporating $\hat{h}_{s_{u}}$, we can calculate the weight coefficient of each POI based on the semantic correlations at different time steps. The formulation for the semantic-user preference attention is as follows.
\begin{equation}
    \begin{split}
        \begin{aligned}
        & Q_S, K_S, V_S = W_{Q_S} \hat{h}_{s_{u}}, W_{K_S} h_{s_{u}}^S , W_{V_S} h_{s_{u}}^S \\
        & \hat{h}_{s_{u}}^S = softmax(\frac{Q_S K_S^T }{\sqrt{d}} )V_S \\ 
        & P_S(u) = \sum_{i=1}^n \hat{h}^S_{s_{u},i}
        \end{aligned}
    \end{split}
    \label{multi-modal}
\end{equation}
where $\hat{h}_{s_{u}}^S$ represents the new semantic representations of the check-in sequence. $P_S(u)$ is the learned semantic POI preference of users, where $\hat{h}^S_{s_{u},i}$ is the $i$-th row in $\hat{h}_{s_{u}}^S$ which is the semantic based representation of node $v_i$ in $s_u$.

The distance-user preference attention is conducted in a similar way. Given the distance-based trajectory representation $h_{s_{u}}^D$ and the personal trajectory representation $\hat{h}_{s_{u}}$,  we can learn the distance-based POI preference $P_D(u)$ according to Eq.(\ref{multi-modal}). %formula ./{multi-modal}
% The distance-user preference attention is conducted in a similar way. First, the distance trajectory representation $h_{s_{u}}^D$ is selected from $h_{G_{D}}$ based on the user sequence $s_{u}$. Then, the personal trajectory representation $\hat{h}_{s_{u}}$ is fused with the distance trajectory representation $h_{s_{u}}^D$ as follows,
% \begin{equation}
%     \begin{split}
%         \begin{aligned}
%         & Q_D, K_D, V_D = W_{Q_D} \hat{h}_{s_{u}}, W_{K_D} h_{s_{u}}^D , W_{K_D} h_{s_{u}}^D \\
%         & \hat{h}_{s_{u}}^D = softmax(\frac{Q_D K_D^T }{\sqrt{d}} )V_D \\ 
%         & P_D(u) = \sum_{i=1}^n \hat{h}^D_{s_{u},i}
%         \end{aligned}
%     \end{split}
%     \label{multi-modal2}
% \end{equation}
% where $\hat{h}_{s_{u}}^D$ represents the new distance representations of the check-in sequence. $P_D(u)$ is the learned distance-based POI preference, where $\hat{h}^D_{s_{u},i}$ is the $i$-th row in $\hat{h}_{s_{u}}^D$ which is the distance based representation of node $v_i$ in $s_u$.

\subsubsection{Multi-view User Preference Fusion}
% 经过上述三个注意层的学习，我们得到了三个偏好
Thanks to the above three attention layers, we successfully get the representation of user preference, i.e., $P(u)$, $P_S(u)$, and $P_D(u)$, from three views. By fully considering these three views of user preference, the probability that a target user $u$ interacts with a candidate POI $v_j$ can be obtained by
\begin{equation}
    \begin{split}
        \begin{aligned}
        z_{u,j} = {P(u)}^T h_{ro,j}  +  {P_D(u)}^T h_{G_D, j} + {P_S(u)}^T h_{G_S, j}
        % P_{ro}(u) = P(u) + P(u)^D + P(u)^T 
        \end{aligned}
    \end{split}
\end{equation}
% where $P(u)$ is user preference learned from section \ref{P(u)}, and $h_{ro,j}$ is the POI representation learned from section \ref{}. $P(u)^D$ and $P(u)^S$ represent the distance and semantic user preference learned from section \ref{}. $h_{G_D, j}$ and $h_{G_S, j}$ donate the POI representations learned from section \ref{}. 
where $z_{u,j}$ is the estimated preference of user $u$ to POI $v_j$. In this way, we can get the recommendation scores over all the candidate POIs $\mathcal{V}$.

\subsection{Objective Function}
\label{objective function} 
We employ a softmax function to calculate the probability of the visit on the next POI from the learned multi-view user preference embedding $z_{u,j}$ as follows,
\begin{equation}
    \begin{split}
        \begin{aligned}
        \hat{y}_{u,j}=\frac{exp(z_{u,j})}{\sum_{v_k \in \mathcal{V}}exp(z_{u,v_k})}.
        \end{aligned}
    \end{split}
    \label{totalloss}
\end{equation}
We employ a cross-entropy loss, as shown in Eq. (\ref{totalloss2}), to optimize the entire trainable parameters $\theta$ with a $\ell_2$ regularization to prevent overfitting.
\begin{equation}
    \begin{split}
        \begin{aligned}
        L = -\sum_{j=1}^N y_{u,j}log(\hat{y}_{u,j})+(1-y_{u,j})log(1-\hat{y}_{u,j})  + \lambda_{\theta} {||\theta||}_2^2
        \end{aligned}
    \end{split}
    \label{totalloss2}
\end{equation}
where $y_{u,j}$ is the ground truth labels in which $y_{u,j}=1$ if $u$'s next visited POI is $v_j$; otherwise, $y_{u,j}=0$. $\lambda_{\theta}$ is a regularization weight.
% $\lambda_{\theta} {||\theta||}^2$ is the L2-regularization for preventing overfitting.

\section{Experiment}
\label{section:5}
% In this section, we will conduct extensive experiments on three real-world LBSN datasets to evaluate the performance of the proposed BayMAN. First, we will describe the experiment setup including the datasets, the way of generating unreliable check-ins, baselines, evaluation metrics, and implementation details. Then we will conduct a quantitative comparison between the proposed BayMAN with the state-of-the-art baselines, followed by a result analysis. Next, the ablation study is conducted to examine whether each component of BayMAN is useful to the robust POI recommendation. Finally, we analyze the effect of the key parameters on the model performance.

\subsection{Experiment Setup}
\textbf{Datasets.} Three LBSN check-in datasets that are widely used in POI recommendations are adopted for evaluation, Gowalla, NYC, and Foursquare. The statistics of the three datasets are shown in Table \ref{DATASETS}. The Gowalla dataset consists of 10,162 users and 24,250 POIs. It contains 307,376 check-in records of these users in California and Nevada over the period of April 2012 to September 2013. The NYC dataset consists of 1,064 users and 5,136 POIs. It contains a total of 147,939 check-in records. The Foursquare dataset consists of 2,321 users and 5,596 POIs. It contains a total of 194,108 check-in records of these users in Singapore over the period of February 2009 to October 2010. We preprocess the dataset and sort the check-in sequence for each user in chronological order. Each dataset is divided into training, validation, and testing sets. For each user, we select the first 80$\%$ of check-in data for training and validation, and the remaining 20$\%$ for testing.

% Brightkite dataset consists of 1,850 users and 1,672 POIs. It contains a total of 40,040 check-ins records of these global users over the period of April 2008 to October 2010. 
% 		Avg.\#check-ins per user&	44.97&	83.63&	140.12\\
%         Avg.\#check-ins per POI	&18.85	&34.69&	155.04\\
\begin{table}
	\small
	\centering
	\caption{Statistics of Three LBSN Datasets}

	\begin{tabular}{c|ccc}
		%\hline
		\hline \toprule[1pt]
		statistic & Gowalla & NYC & Foursquare   \\ 
		\hline 
		\# of user & 10162 & 1064 &2321 \\
		\# of POI&	24238 & 5136	& 	5596\\
		check-in record	&456967	& 147939	&194108\\
        % time range&	2012/04$\sim$2013/09	&2009/02$\sim$2010/10&	2008/04$\sim$2010/10\\
        % cities/countries&	California and Nevada &Singapore&	global\\
		\hline \toprule[1pt]
		%\hline
	\end{tabular}
	\label{DATASETS}
\end{table}

\begin{table*}[]
\renewcommand\arraystretch{1}
\centering
\caption{The performance comparison between BayMAN and the baselines over the three datasets without any data disturbance.}
\small
\begin{tabular}{c|cccc|cccc|cccc}
\hline
          & \multicolumn{4}{c|}{Foursquare}                              & \multicolumn{4}{c|}{NYC}                                     & \multicolumn{4}{c}{Gowalla}                                 \\ \cline{2-13} 
          & R@5 & R@10 & N@5 & \multicolumn{1}{l|}{N@10} & R@5 & R@10 & N@5 & \multicolumn{1}{l|}{N@10} & R@5 & R@10 & N@5 & \multicolumn{1}{l}{N@10} \\ \hline
LSTM      & 0.1501   & 0.2068    & 0.1154 & 0.1369                       & 0.2035   & 0.2649    & 0.1392 & 0.1663                       & 0.1482   & 0.1943    & 0.1096 & 0.1257                      \\
Time-LSTM & 0.1826   & 0.2402    & 0.1347 & 0.1501                       & 0.2209   & 0.2811    & 0.1524 & 0.1812                       & 0.1574   & 0.2193    & 0.1132 & 0.1353                      \\
STGN      & 0.1762   & 0.2374    & 0.1265 & 0.142                        & 0.2437   & 0.3012    & 0.175  & 0.1937                       & 0.1688   & 0.2236    & 0.1205 & 0.1364                      \\
ARNN      & 0.1755   & 0.2381    & 0.1302 & 0.1486                       & 0.1969   & 0.3335    & 0.1362 & 0.2108                       & 0.1537   & 0.2015    & 0.1116 & 0.1282                      \\
LSTPM     & 0.2136   & 0.2703    & 0.1584 & 0.1722                       & 0.2831   & 0.3624    & 0.1941 & 0.2315                       & 0.1798   & 0.2302    & 0.1284 & 0.1452                      \\
ASGNN     & 0.1847   & 0.2528    & 0.1395 & 0.1518                       & 0.2450   & 0.2882    & 0.1763 & 0.1894                       & 0.1836   & 0.2426    & 0.1302 & 0.1503                      \\
STAN      & 0.2715   & 0.3534    & 0.1881 & 0.1969                       & 0.4667   & 0.5863    & 0.3025 & 0.3406                       & 0.1864   & 0.2479    & 0.1327 & 0.152                       \\
GSTN      & 0.2352   & 0.2871    & 0.1683 & 0.1883                       & 0.4276   & 0.5061    & 0.2837 & 0.3129                       & 0.2031   & 0.2637    & 0.1483 & 0.1682                      \\ \hline
BayMAN    & \textbf{0.3039}   & \textbf{0.4118}    & \textbf{0.2049} & \textbf{0.2206}                       & \textbf{0.5392}   & \textbf{0.6667}    & \textbf{0.3254} & \textbf{0.3881}                       & \textbf{0.2364}   & \textbf{0.2981}    & \textbf{0.1698} & \textbf{0.1851}                      \\ \hline
\end{tabular}
\label{normal}
\end{table*}

\textbf{Unreliable Data Generation.} In order to simulate the unreliable data issues including incomplete and noisy data, we adopt three types of data disturbance operations, data deletion, replacement, and a mixture of the two. 
For the data deletion operation, we first set a check-in deletion ratio as $r\%$. Then, $r\%$ percentage of the check-in records in the dataset are randomly selected and removed to simulate the scenario of incomplete check-in data. 
For the replacement operation, we also first set a check-in replacement ratio as $r\%$. Then $r\%$ percentage of the check-in records in the dataset are randomly selected. Instead of directly removing them, we replace them with their distance closest neighbor POIs to simulate the scenario of the noisy data with incorrect check-ins. 
To further test the robustness of the model, we conduct experiments on the data with the mixture of data deletion and replacement operations. Specifically, we set the deletion and replacement ratio as $r\%$. That is, we randomly delete $r\%$ of the check-in records in the data, and also replace $r\%$ of the records.

% {For the operation of the mixture of the two, we set the deletion and replacement ratio to $r\%$.That is, we randomly delete $r\%$ of the check-in records in the data, and also replace $r\%$ of the records}

% For the operation of the mixture of the two, we set the deletion and replacement ratio to $r\%$, randomly select $r\%$ of the check-in records in the data to delete, and $r\%$ of the records to complete the replacement. 
% In the following section, we will conduct experiments under the three types of unreliable data settings

\textbf{Baselines.} We compare BayMAN with the following baselines, containing both classical deep learning models and current state-of-the-art deep learning models for POI recommendation.

\begin{itemize}
    \item \textbf{LSTM} \cite{hochreiter1997long}: LSTM is a variant of the RNN model that is widely used to deal with sequential data. It takes advantage of the sequential dependencies in POI check-in sequences.
    \item \textbf{STGN}\cite{zhao2020go}: STGN equips LSTM with new distance and time gates to consider both spatial and temporal intervals between successive check-ins. It models both short- and long-term POI visiting preferences of users. 
    \item \textbf{ARNN}: ARNN is a state-of-the-art model that uses semantic and spatial information to construct knowledge graphs and improves the performance of the sequential LSTM model.
    \item \textbf{Time-LSTM} \cite{zhu2017next}: Time-LSTM is a state-of-the-art variant of LSTM model for recommendation. It employs two time gates to model the short and long time intervals between continuous check-ins.
    \item \textbf{LSTPM}\cite{wu2019session}: LSTPM is an LSTM-based model that uses a geo-dilated network for capturing the short-term preferences and a non-local network to learn long-term preferences. 
    \item \textbf{ASGNN}\cite{wang2021attentive}: ASGNN is a GNN-based method that employs a personalized hierarchical attention network to capture the long- and short-term preferences of users for the next POI recommendation.
    \item \textbf{STAN}\cite{luo2021stan}: STAN is a state-of-the-art model that uses a bi-attention architecture to learn the spatial and temporal information in non-adjacent locations and non-consecutive check-ins.
    \item \textbf{GSTN} \cite{wanggraph}: GSTN employs a graph-based spatial dependency modeling to capture the high-order geographical influences among POIs, and a LSTM network to capture user-specific temporal preference.
\end{itemize}

\textbf{Evaluation Metric.} We employ two evaluation metrics that are widely used in the task of POI recommendation, recall rate $Recall@K$ and normalized discounted cumulative gain $NDCG@K$.  We recommend the POIs with top-$K$ highest scores and choose the $K$ = \{5,10\} for evaluation. 
% the top-$K$ recall rate $Recall@K$ (R@K) and top-$K$ normalized discounted cumulative gain $NDCG@K$ (N@K) as evaluation metrics, which are widely used in the task of POI recommendation. We recommend the POIs with top-$K$ highest scores and choose the $K$ = \{5,10\} for evaluation.

\textbf {Recall@K} measures the number of the correctly recommended POIs in the top-$k$ recommendation list, which is defined as follows
\begin{equation}
\begin{split}
Recall@K = \frac{\#(hit)}{||D^{test}||},
\end{split}
\end{equation}
where $\#(hit)$ is the number of the target POIs that appear in the top-$k$ recommendation list and $||D^{test}||$ is the size of the test dataset. A higher $Recall@K$ value means a better recommendation result.

\textbf{NDCG@K} measures the quality of the recommendation list, which is defined as follows %ranking lists
\begin{equation}
\small
NDCG@K = 
\left \{ \begin{array}{cc}
\frac{1}{log_2(Rank_i+1)}  &, Rank_i \le K\\
0 &,  Rank_i > K
\end{array}\right.
\end{equation}
where $Rank_i$ represents the position of target POI $v_i$ in the recommendation list. A larger $NDCG@K$ value means a better recommendation result. For brevity, we abbreviate $Recall@K$ and $NDCG@K$ as $R@K$ and $N@K$.

\textbf{Implementation Details.} We set the distance threshold for the distance-based POI graph to 1 $km$, and the replication probability $\varepsilon$ is set to 0.5. The effect of the two hyper-parameters on the model performance will be analyzed in the following section. For the Foursquare, NYC and Gowalla datasets, the number of dimensions for latent representations is set to 48. We utilize Adam optimizer with default betas to optimize the loss function in Section \ref{objective function}. The learning rate is set to 0.002, the number of epochs is 100, and the maximum length for the check-in sequence is set to 50. All the experiments are conducted on a hardware platform with the GPU Nvidia GeForce RTX 3090, 24GB memory, and running python 3.8, cu111, Pytorch 1.8.2. For all the baselines, we set the parameters of the models following the experiment setups in the initial papers.

\begin{table*}[]
\renewcommand\arraystretch{1.17}
\centering
\caption{The performance comparison between BayMAN and baseline methods over the three datasets with the data deletion operation (10\% and 20\% deletion ratios, respectively).}
\begin{tabular}{ccccccccccccc}
\hline
\multicolumn{13}{c}{Check-in data deletion ratio = 10\%}                                                                                                                                                \\ \hline
\multicolumn{1}{c|}{}          & \multicolumn{4}{c|}{Foursquare}                        & \multicolumn{4}{c|}{NYC}                               & \multicolumn{4}{c}{Gowalla}                         \\ \cline{2-13} 
\multicolumn{1}{c|}{}          & R@5    & R@10   & N@5    & \multicolumn{1}{l|}{N@10}   & R@5    & R@10   & N@5    & \multicolumn{1}{l|}{N@10}   & R@5    & R@10   & N@5    & \multicolumn{1}{l}{N@10} \\ \hline
\multicolumn{1}{c|}{LSTM}      & 0.1386 & 0.1852 & 0.1104 & \multicolumn{1}{c|}{0.1311} & 0.1709 & 0.2337 & 0.1252 & \multicolumn{1}{c|}{0.1518} & 0.1316 & 0.1833 & 0.1029 & 0.1192                   \\
\multicolumn{1}{c|}{Time-LSTM} & 0.1636 & 0.2215 & 0.1296 & \multicolumn{1}{c|}{0.1447} & 0.1917 & 0.2584 & 0.1401 & \multicolumn{1}{c|}{0.1703} & 0.1472 & 0.1958 & 0.1084 & 0.1261                   \\
\multicolumn{1}{c|}{STGN}      & 0.1563 & 0.2140 & 0.1203 & \multicolumn{1}{c|}{0.1362} & 0.2048 & 0.2639 & 0.1588 & \multicolumn{1}{c|}{0.1765} & 0.1521 & 0.2004 & 0.1139 & 0.1288                   \\
\multicolumn{1}{c|}{ARNN}      & 0.1507 & 0.2168 & 0.1182 & \multicolumn{1}{c|}{0.1415} & 0.1752 & 0.2813 & 0.1293 & \multicolumn{1}{c|}{0.1904} & 0.1453 & 0.1902 & 0.1063 & 0.1215                   \\
\multicolumn{1}{c|}{LSTPM}     & 0.1895 & 0.2547 & 0.1475 & \multicolumn{1}{c|}{0.1689} & 0.2516 & 0.3239 & 0.1824 & \multicolumn{1}{c|}{0.2172} & 0.1682 & 0.2153 & 0.1206 & 0.1372                   \\
\multicolumn{1}{c|}{ASGNN}     & 0.1631 & 0.2242 & 0.1284 & \multicolumn{1}{c|}{0.1463} & 0.2164 & 0.2608 & 0.1621 & \multicolumn{1}{c|}{0.1753} & 0.1704 & 0.2316 & 0.1241 & 0.1452                   \\
\multicolumn{1}{c|}{STAN}      & 0.2152 & 0.2436 & 0.1651 & \multicolumn{1}{c|}{0.169}  & 0.4073 & 0.4965 & 0.2750 & \multicolumn{1}{c|}{0.3092} & 0.1711 & 0.2198 & 0.1248 & 0.1403                   \\
\multicolumn{1}{c|}{GSTN}      & 0.2073 & 0.2655 & 0.1602 & \multicolumn{1}{c|}{0.1741} & 0.4147 & 0.4943 & 0.2794 & \multicolumn{1}{c|}{0.3081} & 0.1893 & 0.2465 & 0.1366 & 0.1550                   \\ \hline
\multicolumn{1}{c|}{BayMAN}    & \textbf{0.2745} & \textbf{0.3402} & \textbf{0.1916} & \multicolumn{1}{c|}{\textbf{0.2025}} & \textbf{0.5098} & \textbf{0.6275} & \textbf{0.3108} & \multicolumn{1}{c|}{\textbf{0.3620}} & \textbf{0.2052} & \textbf{0.2623} & \textbf{0.1548} & \textbf{0.1699}                   \\ \hline
\multicolumn{13}{c}{Check-in data deletion ratio = 20\%}                                                                                                                                                \\ \hline
\multicolumn{1}{c|}{LSTM}      & 0.1154 & 0.1496 & 0.1001 & \multicolumn{1}{c|}{0.1201} & 0.142  & 0.1944 & 0.1136 & \multicolumn{1}{c|}{0.1382} & 0.1085 & 0.1462 & 0.0913 & 0.1084                   \\
\multicolumn{1}{c|}{Time-LSTM} & 0.1536 & 0.1902 & 0.1246 & \multicolumn{1}{c|}{0.1397} & 0.1796 & 0.2350 & 0.1375 & \multicolumn{1}{c|}{0.1601} & 0.1264 & 0.1725 & 0.0976 & 0.1161                   \\
\multicolumn{1}{c|}{STGN}      & 0.1416 & 0.1802 & 0.1134 & \multicolumn{1}{c|}{0.1226} & 0.1765 & 0.2317 & 0.1490 & \multicolumn{1}{c|}{0.1627} & 0.1286 & 0.1703 & 0.1028 & 0.1153                   \\
\multicolumn{1}{c|}{ARNN}      & 0.1398 & 0.1854 & 0.1127 & \multicolumn{1}{c|}{0.1273} & 0.1531 & 0.2364 & 0.1216 & \multicolumn{1}{c|}{0.1801} & 0.1225 & 0.1671 & 0.0947 & 0.1106                   \\
\multicolumn{1}{c|}{LSTPM}     & 0.1631 & 0.2335 & 0.1343 & \multicolumn{1}{c|}{0.1599} & 0.2243 & 0.2916 & 0.1697 & \multicolumn{1}{c|}{0.2085} & 0.1572 & 0.1943 & 0.1134 & 0.1308                   \\
\multicolumn{1}{c|}{ASGNN}     & 0.1520 & 0.1987 & 0.1226 & \multicolumn{1}{c|}{0.1341} & 0.2065 & 0.2514 & 0.1580 & \multicolumn{1}{c|}{0.1723} & 0.1665 & 0.2128 & 0.1206 & 0.1415                   \\
\multicolumn{1}{c|}{STAN}      & 0.1758 & 0.2261 & 0.1420 & \multicolumn{1}{c|}{0.1603} & 0.3672 & 0.4429 & 0.2563 & \multicolumn{1}{c|}{0.2912} & 0.1493 & 0.1688 & 0.1125 & 0.1247                   \\
\multicolumn{1}{c|}{GSTN}      & 0.2006 & 0.2544 & 0.1518 & \multicolumn{1}{c|}{0.1691} & 0.4025 & 0.4777 & 0.2741 & \multicolumn{1}{c|}{0.3005} & 0.1767 & 0.2294 & 0.1282 & 0.1452                   \\ \hline
\multicolumn{1}{c|}{BayMAN}    & \textbf{0.2679} & \textbf{0.3329} & \textbf{0.1864} & \multicolumn{1}{c|}{\textbf{0.1967}} & \textbf{0.4602} & \textbf{0.5589} & \textbf{0.3020} & \multicolumn{1}{c|}{\textbf{0.3402}} & \textbf{0.1975} & \textbf{0.2479} & \textbf{0.1503} & \textbf{0.1658}                   \\ \hline
\end{tabular}
\label{delete}
\end{table*}

\begin{table*}[]
\renewcommand\arraystretch{1.17}
\centering
\caption{The performance comparison between BayMAN and baseline methods over the three datasets with the data replacement operation (10\% and 20\% replacement ratios, respectively).}
\begin{tabular}{ccccccccccccc}
\hline
\multicolumn{13}{c}{Check-in data replacement ratio = 10\%}                                                                                                                          \\ \hline
\multicolumn{1}{c|}{}          & \multicolumn{4}{c|}{Foursquare}                        & \multicolumn{4}{c|}{NYC}                               & \multicolumn{4}{c}{Gowalla}       \\ \cline{2-13} 
\multicolumn{1}{c|}{}          & R@5    & R@10   & N@5    & \multicolumn{1}{c|}{N@10}   & R@5    & R@10   & N@5    & \multicolumn{1}{c|}{N@10}   & R@5    & R@10   & N@5    & N@10   \\ \hline
\multicolumn{1}{c|}{LSTM}      & 0.1477 & 0.1982 & 0.1126 & \multicolumn{1}{c|}{0.1328} & 0.1895 & 0.2478 & 0.1368 & \multicolumn{1}{c|}{0.1621} & 0.1413 & 0.1886 & 0.1063 & 0.1221 \\
\multicolumn{1}{c|}{Time-LSTM} & 0.1713 & 0.2364 & 0.1305 & \multicolumn{1}{c|}{0.1472} & 0.2027 & 0.2659 & 0.1465 & \multicolumn{1}{c|}{0.1780} & 0.1522 & 0.2101 & 0.1108 & 0.1326 \\
\multicolumn{1}{c|}{STGN}      & 0.1632 & 0.2301 & 0.1221 & \multicolumn{1}{c|}{0.1390} & 0.2248 & 0.2815 & 0.1694 & \multicolumn{1}{c|}{0.1861} & 0.1602 & 0.2154 & 0.1176 & 0.1335 \\
\multicolumn{1}{c|}{ARNN}      & 0.1620 & 0.2312 & 0.1258 & \multicolumn{1}{c|}{0.1445} & 0.1884 & 0.3086 & 0.1362 & \multicolumn{1}{c|}{0.1973} & 0.1498 & 0.1947 & 0.1102 & 0.1268 \\
\multicolumn{1}{c|}{LSTPM}     & 0.2011 & 0.2654 & 0.1524 & \multicolumn{1}{c|}{0.1787} & 0.2652 & 0.3406 & 0.1868 & \multicolumn{1}{c|}{0.2216} & 0.1728 & 0.2216 & 0.1241 & 0.1413 \\
\multicolumn{1}{c|}{ASGNN}     & 0.1735 & 0.2416 & 0.1341 & \multicolumn{1}{c|}{0.1503} & 0.2350 & 0.2749 & 0.1715 & \multicolumn{1}{c|}{0.1832} & 0.1753 & 0.2338 & 0.1275 & 0.1476 \\
\multicolumn{1}{c|}{STAN}      & 0.2439 & 0.2761 & 0.1762 & \multicolumn{1}{c|}{0.1825} & 0.4379 & 0.5521 & 0.2904 & \multicolumn{1}{c|}{0.3228} & 0.1691 & 0.2263 & 0.1228 & 0.145  \\
\multicolumn{1}{c|}{GSTN}      & 0.2325 & 0.2822 & 0.1653 & \multicolumn{1}{c|}{0.1848} & 0.4106 & 0.4887 & 0.2766 & \multicolumn{1}{c|}{0.3041} & 0.2018 & 0.2607 & 0.1454 & 0.1644 \\ \hline
\multicolumn{1}{c|}{BayMAN}    & \textbf{0.2843} & \textbf{0.3431} & \textbf{0.1951} & \multicolumn{1}{c|}{\textbf{0.2075}} & \textbf{0.5196} & \textbf{0.6471} & \textbf{0.3158} & \multicolumn{1}{c|}{\textbf{0.3635}} & \textbf{0.2166} & \textbf{0.2782} & \textbf{0.1603} & \textbf{0.1759} \\ \hline
\multicolumn{13}{c}{Check-in data replacement ratio = 20\%}                                                                                                                          \\ \hline
\multicolumn{1}{c|}{LSTM}      & 0.1310 & 0.1826 & 0.1069 & \multicolumn{1}{c|}{0.1257} & 0.1701 & 0.2213 & 0.1248 & \multicolumn{1}{c|}{0.1496} & 0.1308 & 0.1757 & 0.1012 & 0.117  \\
\multicolumn{1}{c|}{Time-LSTM} & 0.1632 & 0.2148 & 0.1291 & \multicolumn{1}{c|}{0.1412} & 0.1898 & 0.2437 & 0.1395 & \multicolumn{1}{c|}{0.1687} & 0.1496 & 0.2017 & 0.1096 & 0.1315 \\
\multicolumn{1}{c|}{STGN}      & 0.1541 & 0.2143 & 0.1184 & \multicolumn{1}{c|}{0.1365} & 0.2025 & 0.2581 & 0.1564 & \multicolumn{1}{c|}{0.1738} & 0.1561 & 0.2013 & 0.1148 & 0.1302 \\
\multicolumn{1}{c|}{ARNN}      & 0.1538 & 0.2165 & 0.1210 & \multicolumn{1}{c|}{0.1406} & 0.1713 & 0.2724 & 0.1272 & \multicolumn{1}{c|}{0.1850} & 0.1403 & 0.1869 & 0.1073 & 0.1226 \\
\multicolumn{1}{c|}{LSTPM}     & 0.1738 & 0.2394 & 0.1403 & \multicolumn{1}{c|}{0.1655} & 0.2415 & 0.3109 & 0.1754 & \multicolumn{1}{c|}{0.2103} & 0.1629 & 0.2184 & 0.1201 & 0.1385 \\
\multicolumn{1}{c|}{ASGNN}     & 0.1660 & 0.2239 & 0.1301 & \multicolumn{1}{c|}{0.1453} & 0.2246 & 0.2659 & 0.1669 & \multicolumn{1}{c|}{0.1792} & 0.1743 & 0.2245 & 0.1264 & 0.1469 \\
\multicolumn{1}{c|}{STAN}      & 0.2196 & 0.2530 & 0.1662 & \multicolumn{1}{c|}{0.1918} & 0.4103 & 0.4946 & 0.2803 & \multicolumn{1}{c|}{0.3076} & 0.1627 & 0.2057 & 0.1207 & 0.1371 \\
\multicolumn{1}{c|}{GSTN}      & 0.2215 & 0.2741 & 0.1668 & \multicolumn{1}{c|}{0.1759} & 0.4026 & 0.4817 & 0.2726 & \multicolumn{1}{c|}{0.3015} & 0.2002 & 0.2521 & 0.1428 & 0.1615 \\ \hline
\multicolumn{1}{c|}{BayMAN}    & \textbf{0.2705} & \textbf{0.3369} & \textbf{0.1874} & \multicolumn{1}{c|}{\textbf{0.2003}} & \textbf{0.4706} & \textbf{0.6176} & \textbf{0.3075} & \multicolumn{1}{c|}{\textbf{0.3541}} & \textbf{0.2065} & \textbf{0.2643} & \textbf{0.1559} & \textbf{0.1701} \\ \hline
\end{tabular}
\label{replace}
\end{table*}

\begin{table*}[]
\renewcommand\arraystretch{1.17}
\centering
\caption{The performance comparison between BayMAN and baseline methods over the three datasets with the data deletion and replacement operation (10\% and 20\% deletion and replacement ratios, respectively).}
\begin{tabular}{ccccccccccccc}
\hline
\multicolumn{13}{c}{Check-in data deletion and replacement ratio = 10\%}                                                                                                                                                \\ \hline
\multicolumn{1}{c|}{}          & \multicolumn{4}{c|}{Foursquare}                        & \multicolumn{4}{c|}{NYC}                               & \multicolumn{4}{c}{Gowalla}                         \\ \cline{2-13} 
\multicolumn{1}{c|}{}          & R@5    & R@10   & N@5    & \multicolumn{1}{l|}{N@10}   & R@5    & R@10   & N@5    & \multicolumn{1}{l|}{N@10}   & R@5    & R@10   & N@5    & \multicolumn{1}{l}{N@10} \\ \hline
\multicolumn{1}{c|}{LSTM}      & 0.1347 & 0.1840 & 0.1048 & \multicolumn{1}{c|}{0.1250} & 0.1691 & 0.2238 & 0.1229 & \multicolumn{1}{c|}{0.1495} & 0.1225 & 0.1706 & 0.0958 & 0.1114                   \\
\multicolumn{1}{c|}{Time-LSTM} & 0.1601 & 0.2210 & 0.1265 & \multicolumn{1}{c|}{0.1412} & 0.1883 & 0.2492 & 0.1368 & \multicolumn{1}{c|}{0.1647} & 0.1415 & 0.1892 & 0.1024 & 0.1218                   \\

\multicolumn{1}{c|}{STGN}      & 0.1530 & 0.2125 & 0.1172 & \multicolumn{1}{c|}{0.1326} & 0.2010 & 0.2583 & 0.1517 & \multicolumn{1}{c|}{0.1690} & 0.1476 & 0.1914 & 0.1101 & 0.1260                   \\
\multicolumn{1}{c|}{ARNN}      & 0.1488 & 0.2170 & 0.1153 & \multicolumn{1}{c|}{0.1339} & 0.1704 & 0.2759 & 0.1252 & \multicolumn{1}{c|}{0.1861} & 0.1408 & 0.1852 & 0.1022 & 0.1174                   \\
\multicolumn{1}{c|}{LSTPM}     & 0.1792 & 0.2511 & 0.1401 & \multicolumn{1}{c|}{0.1620} & 0.2457 & 0.3114 & 0.1781 & \multicolumn{1}{c|}{0.1904} & 0.1579 & 0.2063 & 0.1135 &  0.1316                  \\
\multicolumn{1}{c|}{ASGNN}     & 0.1570 & 0.2185 & 0.1206 & \multicolumn{1}{c|}{0.1389} & 0.2123 & 0.2591 & 0.1603 & \multicolumn{1}{c|}{0.1728} & 0.1601 & 0.2285 & 0.1176 & 0.1402                  \\
\multicolumn{1}{c|}{STAN}      & 0.2058 & 0.2394 & 0.1563 & \multicolumn{1}{c|}{0.1603}  & 0.3941 & 0.4903 & 0.2698 & \multicolumn{1}{c|}{0.3016} & 0.1642 & 0.2011 & 0.1205 & 0.1383                  \\
\multicolumn{1}{c|}{GSTN}      & 0.2016 & 0.2602 & 0.1520 & \multicolumn{1}{c|}{0.1708} & 0.4101 & 0.4826 & 0.2735 & \multicolumn{1}{c|}{0.3002} & 0.1826 & 0.2407 & 0.1332 & 0.1509                  \\ \hline
\multicolumn{1}{c|}{BayMAN}    & \textbf{0.2647} & \textbf{0.3333} & \textbf{0.1801} & \multicolumn{1}{c|}{\textbf{0.2010}} & \textbf{0.4810} & \textbf{0.6082} & \textbf{0.3040} & \multicolumn{1}{c|}{\textbf{0.3566}} & \textbf{0.2001} & \textbf{0.2542} & \textbf{0.1511} & \textbf{0.1630}                   \\ \hline
\multicolumn{13}{c}{Check-in data deletion and replacement ratio = 20\%}                                                                                                                                                \\ \hline
\multicolumn{1}{c|}{LSTM}    & 0.1046 & 0.1451 & 0.0907 & \multicolumn{1}{c|}{0.1094}  & 0.1342 & 0.1883 & 0.1102 & \multicolumn{1}{c|}{0.1028} & 0.1360  & 0.1431 & 0.0886 & 0.1052                     \\
\multicolumn{1}{c|}{Time-LSTM} & 0.1420 & 0.1896 & 0.1132 & \multicolumn{1}{c|}{0.1263} & 0.1750 & 0.2301 & 0.1315 & \multicolumn{1}{c|}{0.1556} & 0.1213 & 0.1706 & 0.0924 & 0.1148                   \\

\multicolumn{1}{c|}{STGN}      & 0.1322 & 0.1792 & 0.1125 & \multicolumn{1}{c|}{0.1201} & 0.1732 & 0.2285 & 0.1420 & \multicolumn{1}{c|}{0.1572} & 0.1240 & 0.1672 & 0.0988 & 0.1141                   \\
\multicolumn{1}{c|}{ARNN}      & 0.1308 & 0.1805 & 0.1094 & \multicolumn{1}{c|}{0.1258} & 0.1496 & 0.2320 & 0.1183 & \multicolumn{1}{c|}{0.1731} & 0.1179 & 0.1630 & 0.0912 & 0.1091                   \\ 					 		 			
					 				 	
\multicolumn{1}{c|}{LSTPM}     & 0.1520 & 0.2213 & 0.1310 & \multicolumn{1}{c|}{0.1564} & 0.2207 & 0.2901 & 0.1614 & \multicolumn{1}{c|}{0.1972} & 0.1480 & 0.1892 & 0.1054 & 0.1228                   \\
\multicolumn{1}{c|}{ASGNN}     & 0.1491 & 0.1906 & 0.1207 & \multicolumn{1}{c|}{0.1315} & 0.1950 & 0.2449 & 0.1570 & \multicolumn{1}{c|}{0.1683} & 0.1612 & 0.2094 & 0.1150 & 0.1362                   \\
\multicolumn{1}{c|}{STAN}      & 0.1702 & 0.2205 & 0.1368 & \multicolumn{1}{c|}{0.1549}  & 0.3588 & 0.4301 & 0.2512 & \multicolumn{1}{c|}{0.2879} & 0.1395 & 0.1561 & 0.1083 & 0.1197                  \\
\multicolumn{1}{c|}{GSTN}      & 0.1948 & 0.2514 & 0.1499 & \multicolumn{1}{c|}{0.1651} & 0.4016 & 0.4728 & 0.2685 & \multicolumn{1}{c|}{0.2914} & 0.1703 & 0.2215 & 0.1232 & 0.1403                  \\ \hline
\multicolumn{1}{c|}{BayMAN}    & \textbf{0.2549} & \textbf{0.3301} & \textbf{0.1740} & \multicolumn{1}{c|}{\textbf{0.1927}} & \textbf{0.4513} & \textbf{0.5307} & \textbf{0.2910} & \multicolumn{1}{c|}{\textbf{0.3296}} & \textbf{0.1882} & \textbf{0.2350} & \textbf{0.1438} & \textbf{0.1572}                   \\ \hline
\end{tabular}
\label{mixture}
\end{table*}

\subsection{Performance Comparison over Raw Data}
To study whether the proposed BayMAN can achieve better performance over the raw check-in data, we first compare the performance of BayMAN with baselines over the three datasets without any data disturbance operations. The comparison result is shown in Table \ref{normal}. 
The best results are highlighted in bold font.
One can see that BayMAN significantly outperforms all the comparison models, with an 11.93$\%$-16.40$\%$ improvement in recall and a 7.57$\%$-14.50$\%$ increase in NDCG. 
In Foursquare and NYC, compared with the second-ranked STAN, BayMAN improves by 11.93$\%$ and 15.53$\%$ on Recall@5. This is mainly because our method adequately captures the semantic and distance dependencies from a global view, allowing BayMAN to make robust recommendations compared to STAN, which only concentrates on individual user sequences. In Gowalla, BayMAN is 16.40$\%$ higher than the second-ranked model GSTN on Recall@5. It verifies that compared with GSTN which focuses on utilizing global information, we model the semantic dependencies in personal sequences and adopt a novel data augmentation method to cope with the data uncertainty, so as to achieve a robust recommendation. The performance superiority of BayMAN to the baselines implies that the raw POI check-in dataset itself is unreliable and contains noise. 

Among the baseline models, self-attention models such as STAN, ASGNN and GSTN perform better than LSTM-based models, because it is difficult for LSTM-based models to capture the impact of historical check-ins on the next check-in. It should be noted that we do not construct a knowledge graph for ARNN, as the POI category information is not available in our study.
One can also observe that STAN and GSTN perform better than other baselines over the three datasets. STAN focuses more on the personal check-in sequences learning and performs better on Foursquare and NYC, while GSTN learns global geographic information and performs better on Gowalla. This is mainly because the Gowalla dataset is much sparser than the other datasets, and thus it is more difficult to capture each user's preference from the sparse historical check-in sequences. This further validates that the semantic-based and distance-based POI correlations can help better learn the user's preference and result in robust recommendations.

\subsection{Performance Comparison with Incomplete and Noisy POI Check-ins}
To test the robustness of BayMAN on the unreliable POI check-in data, we add the perturbation to the raw datasets through the deletion and replacement operations as described in the experiment setup. We set the ratio of deletion and replacement to 10\% and 20\%, respectively. The results over the three datasets with three types of perturbations are shown in Table \ref{delete}, Table \ref{replace} and Table \ref{mixture}.

% deletion perturbation is shown in Table \ref{delete}, and the result with replacement perturbation is shown in Table \ref{replace}.

As shown in Table \ref{delete}, when 10$\%$ and 20$\%$ of check-in data are randomly removed, the performance of all the methods drops compared with the result shown in Table \ref{normal}. This means the data sparsity issue can truly significantly affect the model performance. In the case of removing 10$\%$ data, BayMAN outperforms all the baseline methods by 6.41$\%$ to 28.14$\%$ over the three datasets in terms of recall rate and improves 9.61$\%$-17.80$\%$ on NDCG. In the case of removing 20$\%$ data, the superiority of BayMAN is more significant than the baselines, with the recall performance improvement ranging from 12.13$\%$ to 34.93$\%$ and NDCG performance improvement 10.18$\%$-22.79$\%$. This shows that BayMAN is more effective than baselines when the data is sparser. From this experiment, one can conclude that BayMAN can achieve more accurate and robust POI recommendation results than baseline methods when users' check-in data are incomplete and sparse.

Table \ref{replace} shows the result of the performance comparison between BayMAN and baseline methods over the datasets with 10\% and 20\% replacement ratios, respectively. It demonstrates that the proposed BayMAN still achieves the best performance in all the cases. Specifically, BayMAN outperforms the baselines by at least 6.71$\%$ and at most 21.58$\%$ on recall, at least 7.10$\%$ and at most 12.61$\%$ on NDCG, when 10$\%$ check-ins are randomly replaced (which can be considered as noise). When the data replacement ratio is increased to 20$\%$, the improvement ranges from 3.15$\%$ to 24.87$\%$ on recall and 5.33$\%$ to 15.12$\%$ on NDCG, depending on different baselines, datasets and top-$k$ values. Similar to the result shown in Table \ref{delete}, the result in Table \ref{replace} also presents the trend that more noise injected into the dataset will lead to more significant performance improvement of BayMAN compared with the baselines. One can also see that compared to the results in Table \ref{normal}, the performance of all the models drops significantly when noise is added, which verifies that incorrect check-ins do affect model performance. This experiment indicates that BayMAN can achieve more accurate and robust POI recommendation results than baseline methods when users' check-in data are noisy with incorrect check-ins.

Table \ref{mixture} shows the performance comparison result under the mixture of deletion and replacement operations with the rations 10\% and 20\%, respectively. BayMAN outperforms all baselines by 5.61\% to 28.62\% in terms of recall and 8.02\% to 18.24\% in terms of NDCG over three datasets when the deletion and replacement ratio is 10\%. When the ratio increases to 20\%, the performance improvement of BayMAN is more significant, achieving recall improvement from 6.09\% to 31.30\% and NDCG improvement from 8.38\% to 16.72\%. This experiment results demonstrate that BayMAN’s POI recommendations are more accurate and robust than those baselines even when the deletion and replacement co-exist in the check-in data.

\begin{figure} \centering
    \subfigure[Foursquare] { 
		\includegraphics[scale=0.265]{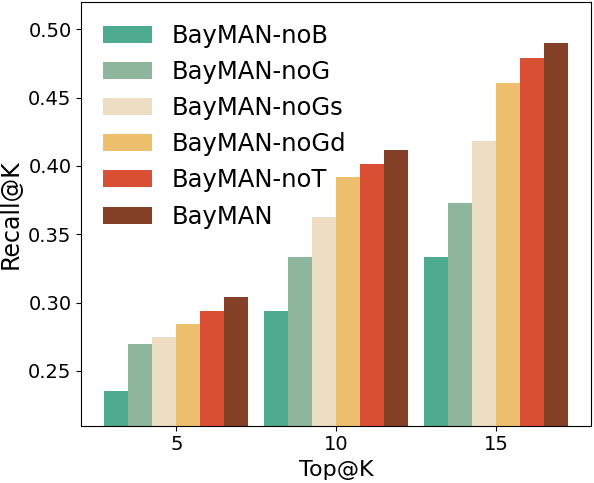}     
	}
	\subfigure[NYC] {
		\includegraphics[scale=0.265]{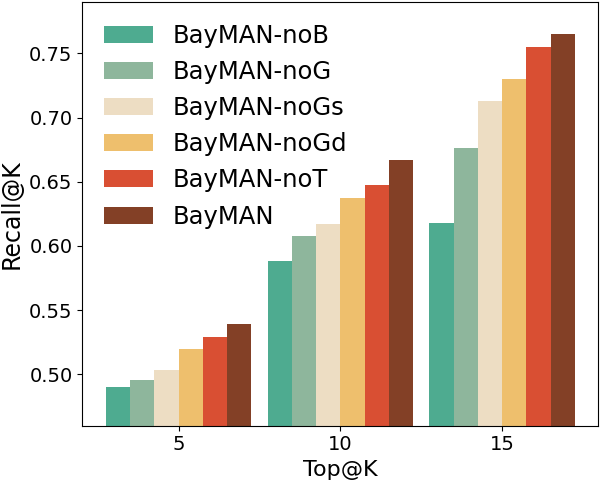} 
	}      
	\caption{Ablation study on Foursquare and NYC dataset.}
	\label{xr} 
\end{figure}

\subsection{Ablation Study}
In this section, we further conduct an ablation study to evaluate the contribution of each component in BayMAN to the performance gain. We deactivate different components and form the following variants.
\begin{itemize}
    \item \textbf{BayMAN-noB}: This variant removes the Bayes-enhanced data augmentation component. That is, the augmented personal POI transition graph $G_u^{\prime}$ is dropped.
    \item \textbf{BayMAN-noG}: This variant removes both the semantic- and the distance-based POI graph. Thus it does not consider the semantic and distance dependencies among POIs. 
    \item \textbf{BayMAN-noGs}: This variant removes the semantic-based POI graph, and thus the semantic dependencies among POIs are ignored.
    \item \textbf{BayMAN-noGd}: This variant removes the distance-based POI graph, and thus the geographic closeness among POIs is ignored.
    \item \textbf{BayMAN-noT}: This variant does not consider the time intervals in the check-in sequence, and removes the time-aware attention layer.
\end{itemize}

Fig. \ref{xr} shows the results of the ablation study. In terms of overall results, we have the following observations. First, BayMAN outperforms its variants in all cases. The significant performance drop of BayMAN-noB compared to BayMAN demonstrates the effectiveness of the Bayes-enhanced data augmentation in capturing the collaborative signals to alleviate the unreliable check-in data. Second, BayMAN-noG exhibits undesirable performance because it does not take into account the semantic relationships and geographic proximity among POIs. This finding highlights the importance of incorporating both semantic and distance information in order to better understand the user preference and more comprehensively explore his/her intention for visiting the next POI. Relying solely on the user preference learned from his/her personal POI transition graph is insufficient for generating robust POI recommendations. Third, the performance of BayMAN-noGd and BayMAN-noGs demonstrates that eliminating either the distance-based POI graph or the semantic-based POI graph leads to a degradation in performance. BayMAN-noGd performs relatively better, possibly because semantic information between POIs can refine the user preference more effectively than distance information in the unreliable user sequence.
Finally, it is evident that BayMAN-noT outperforms all other variants. By modeling both semantic relations and distance dependencies, dealing with uncertainty in personal graphs, and adequately learning user preferences, BayMAN-noT underscores the importance of multi-view graph construction, Bayes-enhanced spatial dependency learning, and multi-view user preference learning. However, despite its superiority over all other variants, BayMAN-noT still lags behind BayMAN, which includes a time-aware attention layer. This suggests that considering time intervals between check-ins is also essential.

\subsection{Parameter Analysis}
\begin{figure} \centering %[!t]
    \subfigure[Deletion] { 
            \includegraphics[scale=0.2]{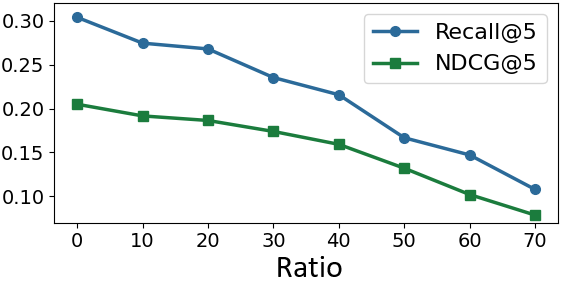}
            \label{range of delete} 
	}
	\subfigure[Replacement] {
            \includegraphics[scale=0.21]{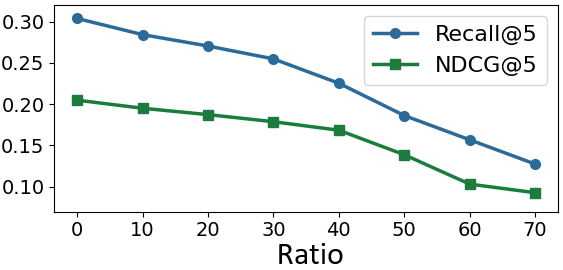}
            \label{range of replace} 
	}      
        \caption{The influence of different data deletion/replacement ratios.} % and
	% \caption{Performance of BayMAN with different ranges of deletion and replacement.}
	\label{delete and replace range} 
\end{figure}

\begin{figure}[!t] \centering %[!t]
    \subfigure[Replacement ratio = 10$\%$] { 
            \includegraphics[scale=0.165]{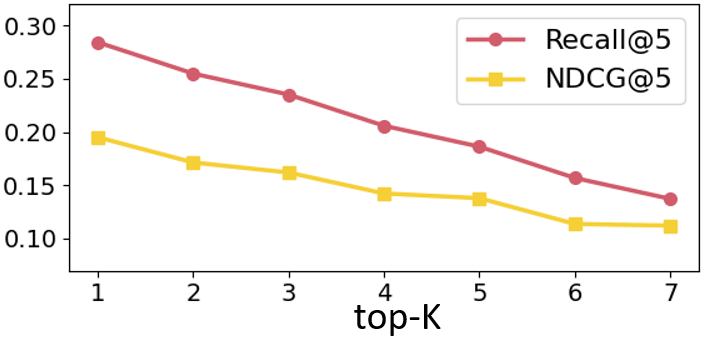}
            \label{replacement ratio 10} 
	}
	\subfigure[Replacement ratio =20$\%$] {
            \includegraphics[scale=0.165]{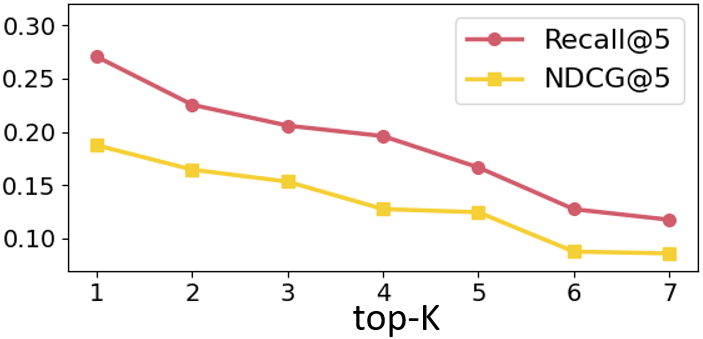}
            \label{replacement ratio 20} 
	}      
        \caption{The influence of parameter K under the top-K replacement strategy.} %
	% \caption{Performance of BayMAN with different ranges of deletion and replacement.}
	\label{topn replace} 
\end{figure}

To investigate the impact of relevant parameters on the unreliable data generation process discussed in Section 5.1, we vary the deletion and replacement ratio $r\%$ and utilize the top-K replacement strategy to evaluate the performance on Foursquare. The results are shown in Fig.\ref{delete and replace range} and Fig.\ref{topn replace}. 
% To investigate the impact of relevant parameters in the unreliable data generation process (discussed in Section 5.1), we vary the deletion and replacement ratio $r\%$ and utilize the top-K replacement strategy to evaluate the proposed BayMAN on Foursquare. The results are shown in Fig. \ref{delete and replace range} and Fig. \ref{topn replace}.

We investigate the impact of deletion and replacement ratios on the next POI recommendation by varying r in \{0, 10\%, 20\%, 30\%, 40\%, 50\%, 60\%, 70\%\}. The results are shown in Fig.\ref{delete and replace range}. For both data deletion and replacement, when r is less than 40\%, the performance of BayMAN on Recall@5 and NDCG@5 presents a slow decreasing trend with the increase of r. However, when r is greater than 40\%, the decline in Recall@5 and NDCG@5 becomes steeper. Therefore, the maximum deletion and replacement ratio that BayMAN can handle is approximately 40\%.

We also test the top-K closest neighbor approach for data replacement by varying K in \{1, 2, 3, 4, 5, 6, 7\} to investigate how it affects the recommendation performance. The results are shown in Fig.\ref{topn replace}. As K increasing, both Recall@5 and NDCG@5 keep decreasing due to the strong spatial correlation among POIs in the close proximity. Selecting a nearby POI for replacement can ensure spatial proximity to the original check-in POI, thereby preserving the similarity between the generated and the original check-ins.

\begin{figure}[!t]
% \begin{figure*}[!t]
\centering

\begin{minipage}[b]{0.47\linewidth}
    \centering
    \includegraphics[width=\linewidth]{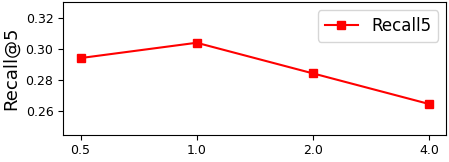}
    \vspace{2pt}
    \includegraphics[width=\linewidth]{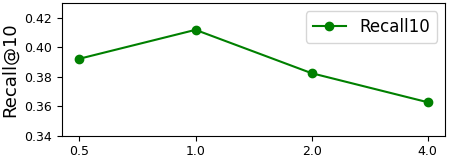}
    \vspace{2pt}
    \includegraphics[width=\linewidth]{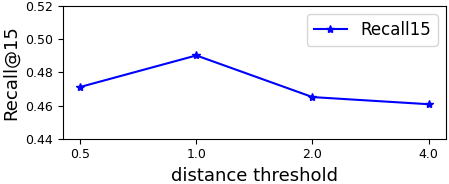}
    \caption*{(a) Foursquare}
    \label{dis:Foursquare}
\end{minipage}
\hfill
\begin{minipage}[b]{0.47\linewidth}
    \centering
    \includegraphics[width=\linewidth]{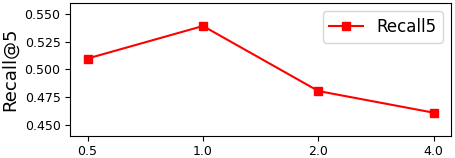}
    \vspace{2pt}
    \includegraphics[width=\linewidth]{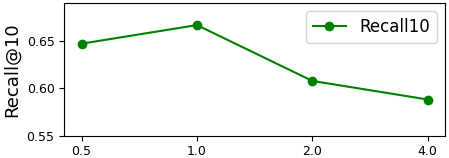}
    \vspace{2pt}
    \includegraphics[width=\linewidth]{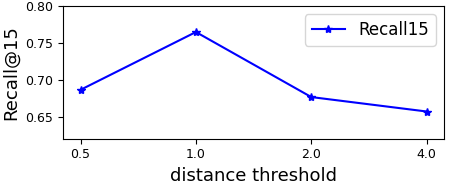}
    \caption*{(b) NYC}
    \label{dis:NYC}
\end{minipage}

\caption{Effect of distance threshold $\Delta d$}
\label{dis}
% \end{figure*}
\end{figure}

\begin{figure}[!t]
% \begin{figure*}[!t]
\centering

\begin{minipage}[b]{0.47\linewidth}
    \centering
    \includegraphics[width=\linewidth]{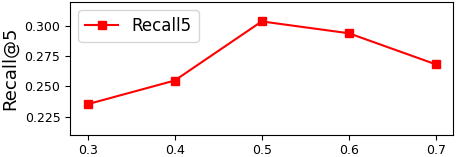}
    \vspace{2pt}
    \includegraphics[width=\linewidth]{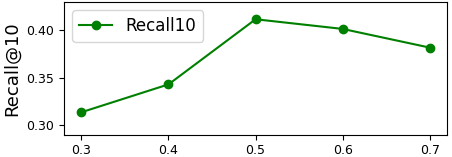}
    \vspace{2pt}
    \includegraphics[width=\linewidth]{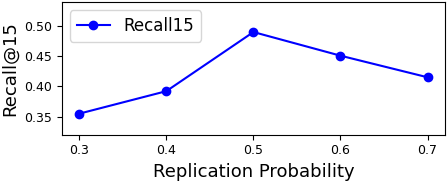}
    \caption*{(a) Foursquare}
    \label{cp:Foursquare}
\end{minipage}
\hfill
\begin{minipage}[b]{0.46\linewidth}
    \centering
    \includegraphics[width=\linewidth]{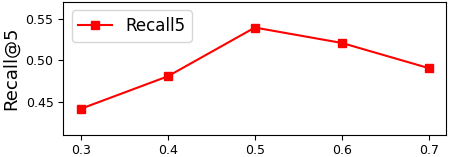}
    \vspace{2pt}
    \includegraphics[width=\linewidth]{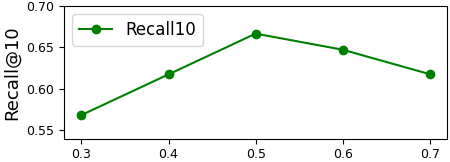}
    \vspace{2pt}
    \includegraphics[width=\linewidth]{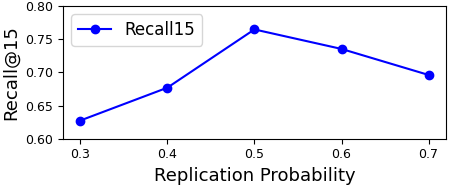}
    \caption*{(b) NYC}
    \label{cp:NYC}
\end{minipage}

\caption{Effect of replication probability $\varepsilon$}
\label{rp}
% \end{figure*}
\end{figure}

To investigate the impact of different key hyperparameters on model performance, we vary the distance threshold $\Delta d$ in Eq. (\ref{Deltad}) and the coefficient $\varepsilon$ in Eq. (\ref{retain}) to evaluate our proposed model on Foursquare and NYC respectively. The results are shown in Fig. \ref{dis} and Fig. \ref{rp}.

To investigate the impact of the distance threshold parameter $\Delta d$ on model performance, we set it to different values and test the performance. Specifically, we vary $\Delta d$ in the distance-based POI graph, setting it to 0.5, 1.0, 2.0, and 4.0 $km$, respectively. The results are shown in Fig. \ref{dis}. Our findings indicate that the best performance is achieved when $\Delta d$ is set to 1.0 $km$. As the distance threshold increases beyond this value, model performance decreases significantly. This is likely due to the weak correlation between distant POIs. This result proves that setting $\Delta d$ to 1.0 $km$ is a suitable choice for capturing the distance correlations.

We next explore the effect of different values of the replication probability $\varepsilon$ on model performance. This hyperparameter determines how many nodes in the original personal graph will be replaced. The higher the node replication probability, the more nodes and the corresponding adjacency will be replaced, and potentially more extra edges and nodes will be gained. Specifically, we vary the value of replication probability $\varepsilon$ from 0.3 to 0.7 and the results are shown in Fig. \ref{rp}. We can observe that the performance improves as the value of the replication probability increases within a certain range. The best performance for both datasets is achieved when $\varepsilon$ is set to 0.5 and then decreased. We attribute the performance degradation of further increasing replication probability value to excessive noises brought by the replacement nodes.

\begin{figure} [!t] %[!t] 
\centering
    \subfigure[Training time] { 
            \includegraphics[scale=0.3]{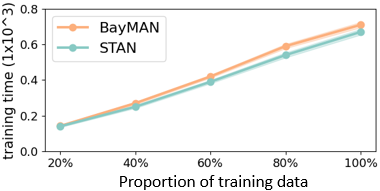}
            \label{time of foursquare} 
	}
    \vspace{2pt}
	\subfigure[Training loss] {
            \includegraphics[scale=0.27]{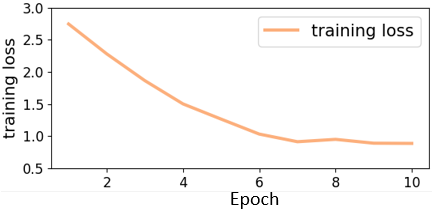}
            \label{time of gowalla} 
	}      
	\caption{Training efficiency. The shadow shows the variance of the training time.} % with varying data proportions
	\label{time} 
\end{figure}

\subsection{Model Efficiency Analysis}

To investigate the time efficiency of BayMAN, we train the model on Foursquare by varying the proportion of training data in \{20\%, 40\%, 60\%, 80\%, 100\%\} and report the average training time for each epoch. The average training time curve for each epoch with the increasing amount of training data is shown in Fig.\ref{time}(a). The average training time of BayMAN for each epoch increases from $1.42\times 10^2$ to $7.11\times 10^2$ seconds, which is very much close to STAN. In addition, the training time of BayMAN shows a linear growth trend when the size of training data scales up. Importantly, BayMAN converges very fast. As the training loss shown in Fig.\ref{time}(b), BayMAN converged in 8 epochs.

% 1111111111111111111111111111111111111111111111111111111

% wom
\subsection{Case Study}
\begin{figure}[t] \centering %[!t]
    \subfigure[Check-in POIs of target and corresponding like-minded user] { 
            \includegraphics[scale=0.187]{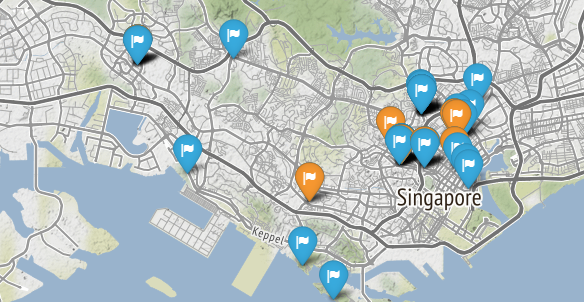} 
            \hspace{0.2cm}  % 或者\quad ，根据需要调整间距大小
            \includegraphics[scale=0.25]{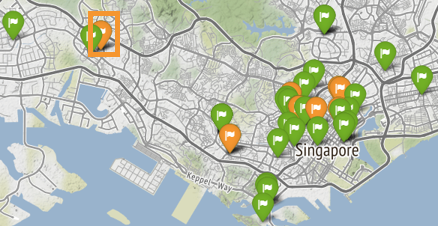}
            \label{check-in POIs} 
	}
	\subfigure[Recommendation results of target user with three methods] {
            \includegraphics[scale=0.24]{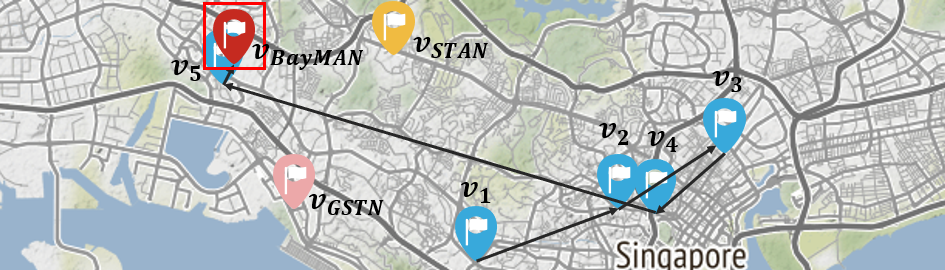}
            \label{check-in sequence} 
	}      
        \caption{A case study on the recommendation performance.} %
	% \caption{Performance of BayMAN with different ranges of deletion and replacement.}
	\label{case study} 
\end{figure}

To further intuitively test the effectiveness of our model, a case study is presented in Fig.\ref{case study}
The left figure in Fig.10(a) visualizes all the POIs visited by a target user $u_{50}$ and the like-minded user $u_{158}$ who shares similar check-in sequences is presented in the right one. Blue, green and orange marks represent the visited POIs by $u_{50}$, $u_{158}$, and both, respectively. The recommendation results for user $u_{50}$ obtained from BayMAN, STAN and GSTN are respectively marked in red, yellow, and pink in Fig.\ref{case study}(b). $u_{50}$’s last 5 check-ins are also shown in Fig.\ref{case study}(b), where arrows indicate the visiting order. The next check-in POI of $u_{50}$, which is the ground truth of the next POI prediction, is marked in a red rectangle. It is obvious that the next POI recommended by BayMAN is much closer to the ground truth than STAN and GSTN, validating the effectiveness of BayMAN in the next POI prediction. BayMAN's ability of learning favorable collaborative signals from the like-minded user $u_{158}$ marked in orange box in Fig.\ref{case study}(a), which enriches and refines the data of the target user.

\section{Conclusion}
\label{section:6}
In this paper, we proposed a Bayesian-enhanced Multi-View Attention Network named BayMAN for robust POI recommendations with unreliable user check-in data. As a unified framework, BayMAN is capable of handling both sparsity and noise in check-in sequences. BayMAN comprehensively modeled the semantic and distance dependencies among POIs by constructing three views of graphs, and adopted a novel data augmentation approach to mitigate the unreliability of the personal check-in sequences. A multi-view attention network was also designed to explore user preferences to POIs refined by the semantic and distance based POI correlations to achieve a robust recommendation. Extensive evaluations on three real datasets verified that the proposed model significantly enhanced the performance and outperformed the state-of-the-art models.

In the future, conducting a deeper study on the unreliable POI check-in data augmentation would be interesting. The data augmentation approach in our model starts from a local view and only utilizes the distance and time information. It is possible to extend the approach to the global view and utilize more extra information. A potential solution is to construct a user-POI bipartite graph using the check-in sequences of all users, and incorporate rich context features, such as POI category and users' social networks. We will consider it as our future work.

\section*{ACKNOWLEDGEMENT}
This research was funded by the National Science Foundation of China (No.62172443), Hunan Provincial Natural Science Foundation of China (No.2022JJ30053), NSF under grant III-2106758, Hong Kong Research Grants Council under Theme-based Research Scheme (No.T41-603/20-R), Hong Kong Jockey Club Charities Trust (No:2021-0369), PolyU RIO (No.BD4A), the Australian Research Council under Future Fellowship (No.FT210100624) and Discovery Project (No. DP190101985).

\ifCLASSOPTIONcaptionsoff
  \newpage
\fi

\normalem
\bibliographystyle{./IEEEtran}
\bibliography{./sigproc}

\vspace{-16mm}
\begin{IEEEbiography}[{\includegraphics[width=1in,height=1.25in,clip,keepaspectratio]{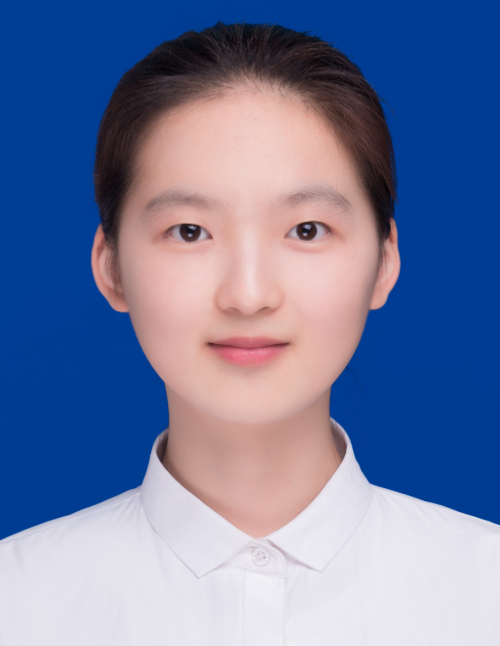}}]{Jiangnan Xia} received the B.E. degree in Software Engineering from NanChang University, Nanchang, China, in 2021. She is currently a Master student of Central South University in the department of Software Engineering. Her research interest includes Spatio-temporal data mining and deep learning.
\end{IEEEbiography}
\vspace{-16mm}
\begin{IEEEbiography}[{\includegraphics[width=1in,height=1.25in,clip,keepaspectratio]{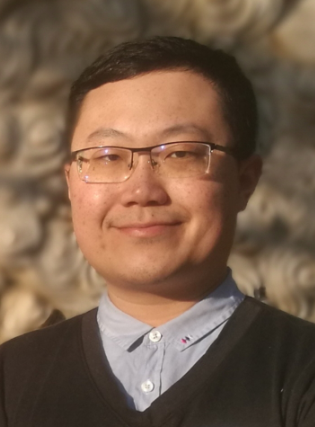}}]{Yu Yang} is currently a Research Assistant Professor with the Department of Computing, The Hong Kong Polytechnic University. He received the Ph.D. degree in Computer Science from The Hong Kong Polytechnic University in 2021. His research interests include spatiotemporal data analysis, representation learning, urban computing, and learning analytics.
\end{IEEEbiography}
\vspace{-16mm}
\begin{IEEEbiography}[{\includegraphics[width=1in,height=1.25in,clip,keepaspectratio]{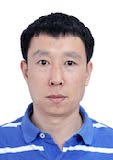}}]{Senzhang Wang}
received the B.Sc. degree from Southeast University, Nanjing, China, in 2009, and the Ph.D. degree in computer science from Beihang University, Beijing, China, in 2015. He is currently a Professor with School of Computer Science and Engineering, Central South University, Changsha. He has published over 100 papers on the top international journals and conferences such as Knowledge and Information Systems, ACM SIGKDD Conference on Knowledge Discovery and Data Mining, AAAI Conference on Artificial Intelligence. His current research interests include data mining and social network analysis.
\end{IEEEbiography}
\vspace{-16mm}
\begin{IEEEbiography}[{\includegraphics[width=1in,height=1.25in,clip,keepaspectratio]{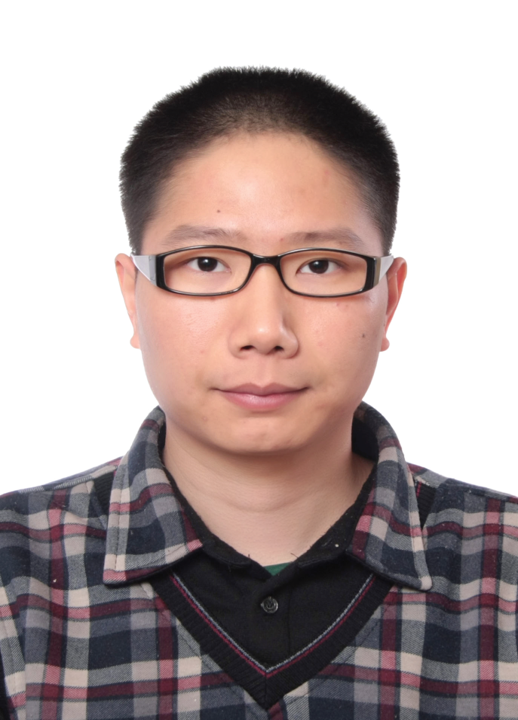}}]{Hongzhi Yin}
received the Ph.D. degree in computer science from Peking University in 2014. He is an Associate Professor with the University of Queensland. He received the Australia Research Council Discovery Early-Career Researcher Award, in 2015. His research interests include recommendation system, user profiling, topic models, deep learning, social media mining, and location-based services.
\end{IEEEbiography}
\vspace{-16mm}
\begin{IEEEbiography}[{\includegraphics[width=1in,height=1.25in,clip,keepaspectratio]{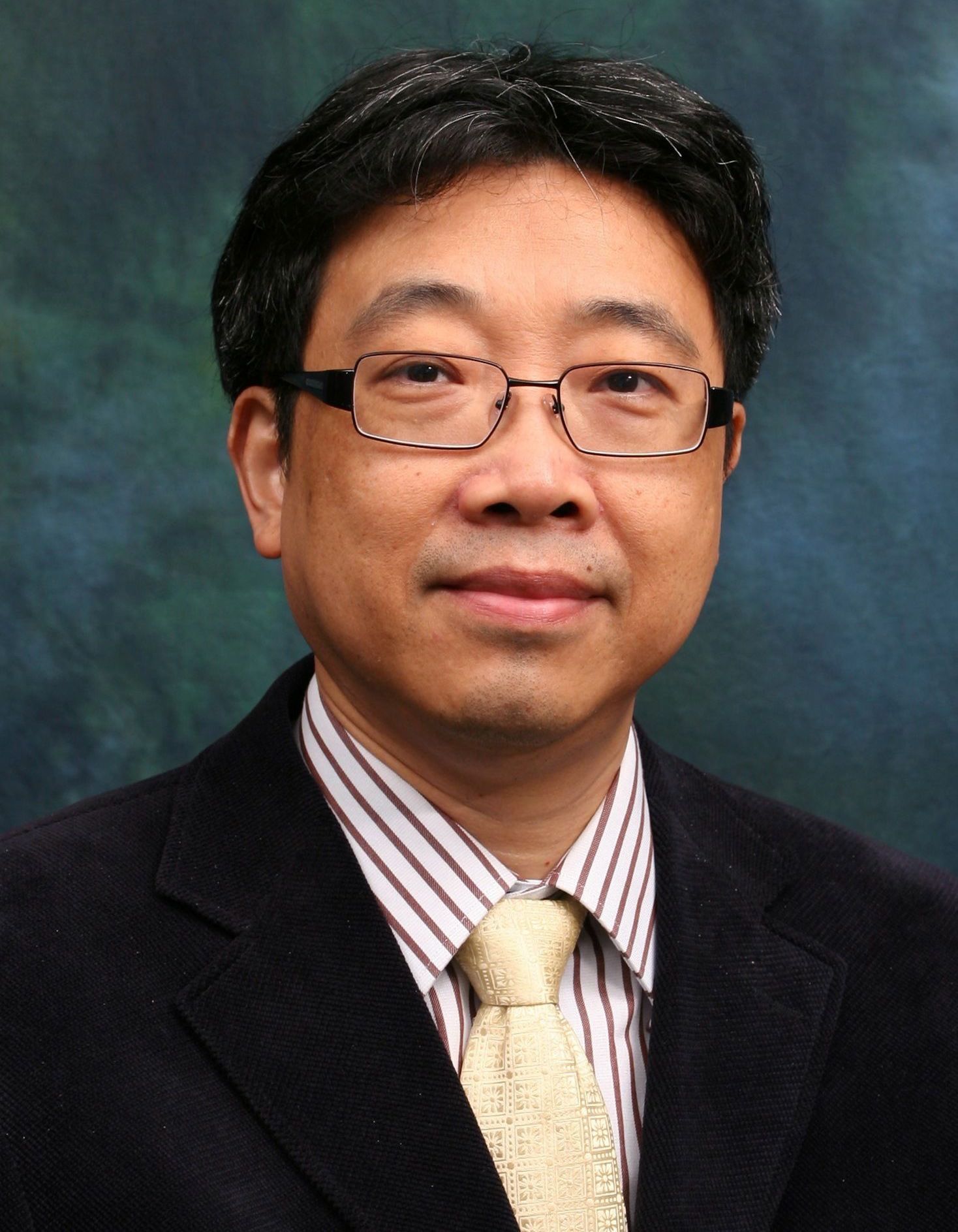}}]{Jiannong Cao} received the M.Sc. and Ph.D. degrees in computer science
from Washington State University, Pullman, WA, USA, in 1986 and 1990, respectively. He is currently the Chair Professor with the Department of Computing, The Hong Kong Polytechnic University, Hong Kong. His current research interests include parallel and distributed computing, mobile computing, and big data analytics. Dr. Cao has served as a member of the Editorial Boards of several international journals, a Reviewer for international journals/conference proceedings, and also as an Organizing/ Program Committee member for many international conferences.
\end{IEEEbiography}
\vspace{-16mm}
\begin{IEEEbiography}[{\includegraphics[width=1in,height=1.25in,clip,keepaspectratio]{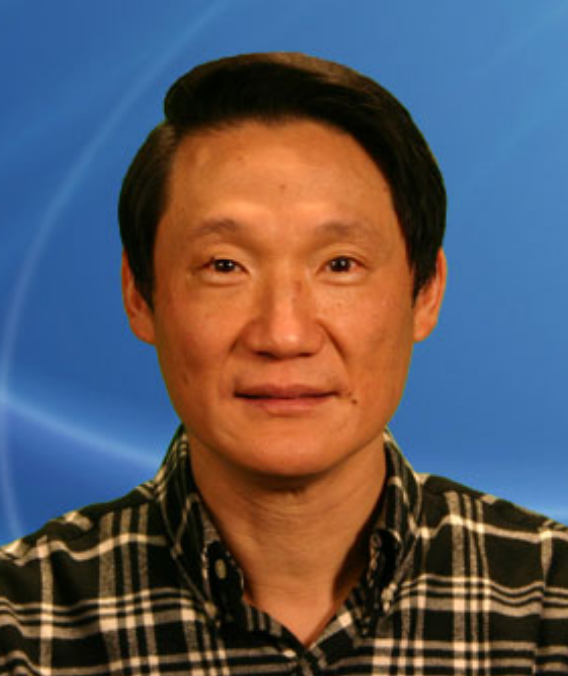}}]{Philip S. Yu} received the B.S. Degree in E.E. from National Taiwan University, the M.S. and Ph.D. degrees in E.E. from Stanford University, and the M.B.A. degree from New York University. He is a Distinguished Professor in Computer Science at the University of Illinois at Chicago and also holds the Wexler Chair in Information Technology. Before joining UIC, Dr. Yu was with IBM, where he was manager of the Software Tools and Techniques department at the Watson Research Center. His research interest is on big data, including data mining, data stream, database and privacy. He has published more than 1,100 papers in refereed journals and conferences. He holds or has applied for more than 300 US patents. Dr. Yu is a Fellow of the ACM and the IEEE.
\end{IEEEbiography}
% \vspace{-16mm}
% \begin{IEEEbiography}[{\includegraphics[width=1in,height=1.25in,clip,keepaspectratio]{image/Senzhang_Wang}}]{Senzhang Wang}
% received the B.Sc. degree from Southeast University, Nanjing, China, in 2009, and the Ph.D. degree in computer science from Beihang University, Beijing, China, in 2015. He is currently a Professor with School of Computer Science and Engineering, Central South University, Changsha. He has published over 100 papers on the top international journals and conferences such as Knowledge and Information Systems, ACM SIGKDD Conference on Knowledge Discovery and Data Mining, AAAI Conference on Artificial Intelligence. His current research interests include data mining and social network analysis.
% \end{IEEEbiography}

% Philip S. Yu (Fellow, IEEE) received the BS degree in electrical engineering from National Taiwan University, the MS and PhD degrees in electrical engineering from Stanford University, and the MBA. degree from New York University. He is a distinguished professor in computer science with the University of Illinois at Chicago and also holds the Wexler Chair in Information Technology. Before joining UIC, he was with IBM, where he was manager of the Software Tools and Techniques Department, Watson Research Center. His research interests include big data, including data mining, data stream, database, and privacy. He has published more than 1,100 papers in refereed journals and conferences. He holds or has applied for more than 300 US patents. He is a fellow of the ACM.
\end{document}